%% file: DMS_20250315.tex
\documentclass[aps,twocolumn,amsfonts,showpacs,pra,amsmath,amssymb,superscriptaddress]{revtex4-1}
\usepackage{graphicx, color}
\usepackage{hyperref}
\hypersetup{colorlinks=true,linkcolor=blue,citecolor=blue,urlcolor=blue}
\usepackage{dcolumn}
\usepackage{bm}
\usepackage{mathtools}
\usepackage{newtxtext}
\usepackage{newtxmath}
\usepackage{anyfontsize}
\usepackage[version=3]{mhchem}
\usepackage[normalem]{ulem}
\usepackage{lipsum}
\usepackage{braket}
\usepackage{comment}

\DeclareMathOperator{\argmin}{\mathrm{argmin}}

\DeclareMathOperator{\divergence}{div}
\DeclareMathOperator{\sinc}{sinc}
\let\Re\relax
\DeclareMathOperator{\Re}{Re}
\let\Im\relax
\DeclareMathOperator{\Im}{Im}
\DeclareMathOperator{\spanspace}{span}
\DeclareMathOperator{\Tr}{Tr}
\DeclareMathOperator{\sgn}{sgn}

\begin{document}

\title{Connecting Density Matrix Spectroscopy to Biexciton Entanglement Dynamics}

\newcommand{\TohokuUniv}{Department of Applied Physics, Tohoku University, Sendai 980-8579, Japan}

\author{Yusuke Masaki}
\affiliation{\TohokuUniv}
\affiliation{Research and Education Center for Natural Sciences, Keio University, Hiyoshi 4-1-1, Yokohama, Kanagawa 223-8521, Japan}

\author{Takashi Otaki}
\altaffiliation{Present address: Nippon Steel Corporation, 20-1 Shintomi, Futtsu, Chiba 293-8511, Japan}
\affiliation{\TohokuUniv}

\author{Hiroaki Matsueda}
\affiliation{\TohokuUniv}
\affiliation{Center for Science and Innovation in Spintronics, Tohoku University, Sendai 980-8577, Japan}
\date{\today}

\input{DMS_abstract}
\maketitle

\input{DMS_Introduction}
\input{DMS_SecII-A}
\input{DMS_SecII-B}
\input{DMS_SecII-C}

\input{DMS_SecIII}
\input{DMS_SecIV-A}
\input{DMS_SecIV-B}
\input{DMS_SecIV-C}

\input{DMS_SecIV-D}
\input{DMS_SecIV-E-1}

\input{DMS_SecIV-E-2}

\input{DMS_SecIV-E-3}
\input{DMS_Summary}

\appendix
\input{DMS_AppendixA}
\input{DMS_AppendixB}
\input{DMS_AppendixC}

\input{DMS_AppendixD}

\bibliographystyle{apsrev4-1}

\input{DMS_References}
\end{document}

%% file: DMS_abstract.tex
\begin{abstract}
  Quantum entanglement is one of the most intriguing features of quantum mechanics.
  To investigate the entanglement between two excitons in a biexciton, an experimental technique called density matrix spectroscopy (DMS) has recently been developed.
  DMS combines stimulated emission tomography and pump-probe techniques to obtain a time-resolved density matrix of the polarization state of a photon pair emitted from the biexciton. 
  The reconstructed density matrix is expected to encode information about the biexciton state and its entanglement dynamics, but the precise nature of this connection has remained unclear.
  In this paper, we derive an analytical relationship between the density matrix obtained by DMS and the biexciton state.
  In addition, we perform numerical simulations to compare the entanglement dynamics obtained by DMS with the biexciton's entanglement dynamics in a two-dimensional electron-hole system using an extended ionic Hubbard model. 
  We find that DMS can partially capture the entanglement in the biexciton, in particular, the dynamics of the difference $S_{\mathrm{bi}} - S_k$, where $S_{\mathrm{bi}}$ is the entanglement entropy of the biexciton and $S_k$ is the entanglement in terms of the wavevectors of the excitons that constitute the biexciton. These results demonstrate the validity of DMS for obtaining information about the entanglement dynamics of the biexciton.
\end{abstract}

%% file: DMS_Introduction.tex
\section{\label{SecI}Introduction}
\begin{figure}[b]
\includegraphics[width=6cm]{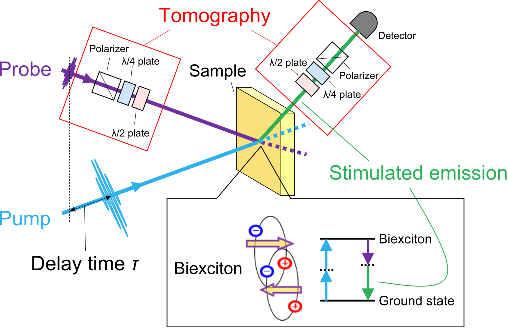}
\caption{\label{fig:DMS} Schematic of density matrix spectroscopy.}
\end{figure}

Quantum entanglement is a key feature not only in quantum information technology~\cite{Nielsen} but also in condensed matter physics~\cite{Amico}.
Electrons or spins in matter are typical quantum many-body systems, and entanglement offers a powerful framework to characterize many-electron or many-spin states~\cite{Amico,Osterloh,Kitaev,Jiang,Walsh,Shim,Gioev,Haldane}.
Although entanglement in quantum materials is difficult to investigate in general due to the lack of a universal probing method, various detection approaches have been reported~\cite{Liu,Pichler,Islam,Cramer,Hales,Scheie1,Scheie2,Lin,Sifain,Rappoport,Garlatti,Fukuhara,Ranni,Brydges}.
Here, we focus on entanglement dynamics in a biexciton, which provides a simple example of a quantum many-body system in condensed matter physics.
The biexciton is widely known as a source of entangled photon pairs~\cite{Edamatsu,Oohata_PRL}.
In semiconductors such as CuCl, a biexciton is converted into a pair of photons with polarization entanglement, which reflects the entanglement between the two excitons forming the biexciton.
It is expected that some information about the entanglement of the biexciton can be extracted by analyzing the polarization state of photons emitted from the biexciton.
To measure the time-resolved density matrix $\hat{\rho}_{\mathrm{DMS}}$ of the polarization state of the photon pair, density matrix spectroscopy (DMS) has been developed~\cite{Oohata_DMS}.
This is achieved by combining stimulated emission tomography and pump-probe techniques as shown in Fig.~\ref{fig:DMS}.
In DMS, a pump pulse and a probe pulse separated by a time delay $\tau$ are incident on the sample, and a polarization-resolved four-wave mixing (FWM) signal is measured.
The pump pulse is tuned to resonantly excite a biexciton through two-photon absorption.
When the probe pulse interacts with the biexciton, an FWM signal is generated.
The density matrix $\hat{\rho}_{\mathrm{DMS}}(\tau)$ can be reconstructed from the polarization-resolved FWM signal obtained for various probe-pulse polarizations~\cite{SET}. Although the entanglement dynamics evaluated from $\hat{\rho}_{\mathrm{DMS}}(\tau)$ is expected to reflect the dynamics of the biexciton, 
the lack of a theoretical description of DMS leaves the connection between $\hat{\rho}_{\mathrm{DMS}}(\tau)$ and the biexciton state unclear, particularly in terms of their entanglement dynamics. In this study, we establish the theoretical framework for DMS and clarify what information about the entanglement of the biexciton can be extracted from DMS.

This paper is organized as follows.
In Sec.~\ref{SecII} and Sec.~\ref{SecIII}, we formulate DMS and examine the relation between $\hat{\rho}_{\mathrm{DMS}}(\tau)$ and the biexciton state.
To properly incorporate phase-matching conditions of FWM, we divide the discussion into two cases based on the sample thickness. The thickness of the sample considered in Sec.~\ref{SecII} (\ref{SecIII}) is much greater (smaller) than the wavelength of the light pulses. 
In Sec.~\ref{SecIV}, we perform numerical simulations of DMS for a two-dimensional electron-hole system to investigate the relationship between the entanglement of the biexciton and that encoded in $\hat{\rho}_{\mathrm{DMS}}$.
We use an extended ionic Hubbard model as the electron-hole system.
We compare the entanglement dynamics in the biexciton and that obtained by DMS, and examine the effectiveness of DMS for investigating the biexciton entanglement dynamics. A summary is given in Sec.~\ref{SecV}. 
Derivations of several formulae and the multiple-reflection effects on the FWM signal are presented in Appendices~\ref{AppendixA}, \ref{AppendixB}, \ref{AppendixC}, and \ref{AppendixD}.

%% file: DMS_SecII-A.tex
\section{\label{SecII}Formulations for thick samples}
\subsection{\label{SecIIA}Derivation of FWM light}
In this section, we formulate DMS for the samples whose thickness $L_z$ is sufficiently large compared with the wavelength of the light ($L_z \gg \lambda$).
Because stimulated emission can be safely described with classical light fields, we combine Maxwell's equations with the Schr\"{o}dinger equation for an electron-hole system.
We assume that the sample is an isotropic three-dimensional, uncharged insulating crystal with the refractive index $n(\omega)$.
It occupies the region $-L_x/2 < x < L_x/2$, $-L_y/2<y<L_y/2$, and $0<z<L_z$, where $L_z$ is the sample thickness, while $L_x$ and $L_y$ are taken to be infinitely large for simplicity.
We assume both the incident pump and probe pulses come from $z<0$, and the transmitted FWM light is detected at $z>L_z$.
Multiple reflections are ignored for simplicity in this section.
Coulomb gauge ($\divergence {\bf A} = 0$) is used throughout this paper.

First, we introduce the transverse fields and the notation of their linear polarization in advance.
A vector field with a wave vector ${\bf k}=(k_x,k_y,k_z)$, written as ${\bf v}_{{\bf k}}e^{i{\bf k}\cdot {\bf r}}$, can be decomposed as $({\bf v}_{\mathrm{T},{\bf k}} + {\bf v}_{\mathrm{L},{\bf k}} )e^{i{\bf k}\cdot {\bf r}}$, where ${\bf k}\cdot {\bf v}_{\mathrm{T},{\bf k}} = 0$ and ${\bf k}\times {\bf v}_{\mathrm{L},{\bf k}} = {\bf 0}$. The subscript $\mathrm{T}$ ($\mathrm{L}$) denotes the transverse (longitudinal) field component.
Explicitly, the components are given by ${\bf v}_{\mathrm{T},{\bf k}} =  {\bf v}_{{\bf k}}- \left({\bf e}_{\bf k}\cdot {\bf v}_{{\bf k}}\right){\bf e}_{\bf k}$ and ${\bf v}_{\mathrm{L},{\bf k}} =   \left({\bf e}_{\bf k}\cdot {\bf v}_{{\bf k}}\right){\bf e}_{\bf k}$ with ${\bf e}_{\bf k} \coloneq {\bf k}/|{\bf k}|$. The transverse component can further be decomposed as ${\bf v}_{\mathrm{T},{\bf k}} = \sum_{\alpha = s,p}v_{\mathrm{T},{\bf k}}^\alpha  {\bf e}_{{\bf k}}^\alpha$ (${\bf e}_{{\bf k}}^\alpha \in \mathbb{R}^3$), where $\{ {\bf e}_{\bf k}, {\bf e}_{{\bf k}}^s, {\bf e}_{{\bf k}}^p \}$ is an orthonormal set and ${\bf e}_{{\bf k}}^s \perp {\bf e}_{z}$ with ${\bf e}_z$ the unit vector along the $z$ direction.
$v_{\mathrm{T},{\bf k}}^\alpha$ is referred to as the $\alpha$-polarized component of ${\bf v}_{\mathrm{T},{\bf k}}$.

The FWM light is generated by the transverse component of the nonlinear current density produced via the FWM process, ${\bf J}_{\mathrm{T,FWM}}$, which is derived from Maxwell's equations.
We define ${\bf J}_{\mathrm{T},\mathrm{FWM},{\bf k}}$ as the Fourier component of ${\bf J}_{\mathrm{T,FWM}}$ (${\bf J}_{\mathrm{T,FWM}} =\sum_{{\bf k}} {\bf J}_{\mathrm{T},\mathrm{FWM},{\bf k}} e^{i{\bf k}\cdot {\bf r}}$, $k_i = 2\pi n_i/L_i$, $n_i \in \mathbb{Z}$).
The electric field of the transmitted FWM light at position ${\bf r} = (x,y,z)$ and angular frequency $\omega$, denoted by ${\bf E}_{\mathrm{FWM}}({\bf r},\omega)$, takes the form
\begin{align}
  {\bf E}_{\mathrm{FWM}}({\bf r},\omega) &= \sum_{{\bf k}_{\parallel}} {\bf E}_{\mathrm{FWM},{\bf k}_{\parallel}}(\omega)
  e^{ i {\bf k}_{\mathrm{tr}}\cdot {\bf r} },   \\
   E_{\mathrm{FWM},{\bf k}_{\parallel}}^\alpha (\omega)&= t_{\mathrm{sv},{\bf k}_0}^\alpha
  \left[\sum_{k_z}  f_{{\bf k}} (\omega) J_{\mathrm{T},\mathrm{FWM},{\bf k}}^\alpha (\omega)
    \right], \label{eq:E_FWM} \\
   f_{\bf k}(\omega)&= -\frac{\mu_0 \omega}{2k_0} L_z \sinc\left(\frac{\Delta k}{2}L_z \right)
  e^{ i(k_0-k_{\mathrm{tr}}- \Delta k/2)L_z}.
  \label{eq:f_k_def}
\end{align}
Here, ${\bf k}_{\mathrm{tr}} = (k_x, k_y, k_{\mathrm{tr}})$ with $k_{\mathrm{tr}} \coloneq \sqrt{\omega^2/c^2-|{\bf k}_{\parallel}|^2}$ and ${\bf k}_{\parallel} = k_x {\bf e}_x + k_y {\bf e}_{y}$, where ${\bf e}_x$ (${\bf e}_y$) is a unit vector along the $x$ ($y$)-axis. 
A polarization-independent factor $f_{\bf k}$ arises from the propagation of light with ${\bf k} = (k_x, k_y, k_z)$. The coefficient $t_{\mathrm{sv},{\bf k}_{0}}^\alpha$ ($\alpha = s$, $p$) denotes the amplitude transmittance of $\alpha$-polarized light with the wavevector ${\bf k}_0 = \left( k_x,k_y,k_0 \right)$ from the sample to vacuum, where $k_0\coloneq \sqrt{n(\omega)^2\omega^2/c^2-|{\bf k}_{\parallel}|^2}$. 
The phase mismatch is defined as $\Delta k \coloneq k_0 - k_z$.
The derivation of Eq.~(\ref{eq:E_FWM}) is found in Appendix~\ref{AppendixA}.
Note that we adopt the following sign convention and normalization for the Fourier transform and its inverse:
\begin{align}
&F_{\bf k}(\omega) = \frac{1}{2\pi L_x L_y L_z} \int_{\mathrm{Sample}}\mathrm{d}{\bf r}\, \int^{\infty}_{-\infty} \mathrm{d}t \, 
\, F({\bf r},t) e^{-i{\bf k}\cdot {\bf r} + i\omega t}, \nonumber \\
&F({\bf r},t) = \sum_{\bf k} \int^{\infty}_{-\infty} \mathrm{d}\omega \, F_{{\bf k}}(\omega) e^{i{\bf k}\cdot {\bf r} - i\omega t} .
\label{eq:FT-def}
\end{align}

To discuss the relation between the FWM signal and the biexciton wavefunctions, an analytical representation of ${\bf J}_{\mathrm{T},\mathrm{FWM},{\bf k}}(\omega)$ ($\omega > 0$) is obtained by a perturbative approach.
The pump pulse is applied to the sample at time $t=0$, exciting biexcitons, and we focus on the dynamics for $t > 0$.
In this time interval, we consider the semiclassical light-matter Hamiltonian
\begin{align}
  \hat{H}(t,\tau) &= \hat{H}_0 - \int \mathrm{d}{\bf r} \, \hat{{\bf J}}_{\mathrm{T}}({\bf r}) \cdot {\bf A}_{\mathrm{pr}}({\bf r},t,\tau) \nonumber  \\
  &= \hat{H}_0 - V\sum_{{\bf k}} \hat{{\bf J}}_{\mathrm{T},{-\bf k}}\cdot {\bf A}_{\mathrm{pr},{\bf k}}(t,\tau),
  \label{eq:Hamiltonian}
\end{align}
where $\hat{H}_0$ is the matter part of the Hamiltonian, $\hat{{\bf J}}_{\mathrm{T}}$ is the transverse current density operator, ${\bf A}_{\mathrm{pr}}$ is the vector potential of the probe pulse in the sample, and $V$ is the volume of the sample: $V\coloneq L_xL_yL_z$.
Here, $\tau$ represents the delay time, corresponding to the arrival of the probe pulse at $t = \tau$.
Periodic boundary conditions are imposed for the quantum mechanical calculation.
In this paper, we only consider pure states as the state of the sample.
The state at time $t$ is denoted by $\ket{\psi(t, \tau)}$.
The state at $t=t_0$ $(t_0<\tau)$ is denoted by $\ket{\psi(t_0,\tau)} \equiv \ket{\psi_0}$.
The state $\ket{\psi(t, \tau)}$ evolves in time $t$ according to the Schr\"{o}dinger equation
$i\hbar \partial_t \ket{\psi(t,\tau)} = \hat{H}(t,\tau) \ket{\psi(t,\tau)}$.
The expectation value of the transverse current density takes the form
\begin{align}
  \braket{\psi(t,\tau)|\hat{{\bf J}}_{\mathrm{T},{\bf k}}|\psi(t,\tau)} =
  &{\bf J}_{\mathrm{pu},\mathrm{T},{\bf k}}(t) + \Delta {\bf J}_{\mathrm{T},{\bf k}} (t,\tau) \nonumber \\
  &+ O \left(A_{\mathrm{pr}}^2 \right),
\end{align}
where ${\bf J}_{\mathrm{pu},\mathrm{T},{\bf k}}(t)$ is independent of ${\bf A}_{\mathrm{pr}}$, and $\Delta{\bf J}_{\mathrm{T},{\bf k}}$ denotes the part of the transverse current proportional to ${\bf A}_{\mathrm{pr}}$.
The transverse FWM current ${\bf J}_{\mathrm{T},\mathrm{FWM},{\bf k}}$ appearing in Eq.~(\ref{eq:E_FWM}) is the FWM component of $\Delta{\bf J}_{\mathrm{T},{\bf k}}$, as specified below.
By applying the time-dependent perturbation theory, we can obtain
\begin{widetext}
  \begin{align}
    \Delta J_{\mathrm{T},{\bf k}}^\alpha (\omega,\tau) &=
    V\sum_{\beta = s,p}\sum_{nml}\sum_{{\bf k}'}
    \left[
      \frac{\braket{\psi_0|n}  \braket{n|\hat{J}^\beta_{\mathrm{T},-{\bf k}'}|l}  \braket{l|\hat{J}^\alpha_{\mathrm{T},{\bf k}}|m}  \braket{m|\psi_0}}{\hbar \omega + \epsilon_l - \epsilon_m + i\eta}
      -\frac{\braket{\psi_0|n}  \braket{n|\hat{J}^\alpha_{\mathrm{T},{\bf k}}|l}  \braket{l|\hat{J}^\beta_{\mathrm{T},-{\bf k}'}|m}  \braket{m|\psi_0}}{\hbar \omega + \epsilon_n - \epsilon_l + i\eta}
      \right] \nonumber \\
    &\hspace{1em}\times \exp \left( -i\frac{\epsilon_n - \epsilon_m}{\hbar} t_0 \right) A_{\mathrm{pr},{\bf k}'}^\beta \left( \omega + \frac{\epsilon_n - \epsilon_m}{\hbar}, \tau \right),
    \label{eq:DeltaJ_psi0}
  \end{align}
\end{widetext}
where 
$\ket{n}$, $\ket{m}$, and $\ket{l}$ are eigenstates of $\hat{H}_0$, with eigenenergies $\epsilon_n$, $\epsilon_m$, and $\epsilon_l$, respectively, where the ordering is $\epsilon_0 < \epsilon_1 \leq \epsilon_2 \leq \cdots$, and $\eta$ is an infinitesimal positive energy.
We assume that the shape of the probe vector potential does not depend on the delay time $\tau$, which is expressed as ${\bf A}_{\mathrm{pr},{\bf k}}(t,\tau) =  {\bf A}_{\mathrm{pr},{\bf k}}(t-\tau,0)$.
Using the Fourier transform, we can obtain
\begin{equation}
  {\bf A}_{\mathrm{pr},{\bf k}}(\omega,\tau) =  {\bf A}_{\mathrm{pr},{\bf k}}(\omega,0)\exp(i\omega \tau).
  \label{eq:Apr_condition}
\end{equation}
By substituting this relation into Eq.~(\ref{eq:DeltaJ_psi0}), we can derive
\begin{widetext}
  \begin{align}
    \Delta J_{\mathrm{T},{\bf k}}^\alpha (\omega,\tau) &=
    V\sum_{\beta = s,p}\sum_{nml} \sum_{{\bf k}'}
    \left[
      \frac{\braket{\psi_{\mathrm{pu}}(\tau)|n}  \braket{n|\hat{J}^\beta_{\mathrm{T},-{\bf k}'}|l}  \braket{l|\hat{J}^\alpha_{\mathrm{T},{\bf k}}|m}  \braket{m|\psi_{\mathrm{pu}}(\tau)}}{\hbar \omega + \epsilon_l - \epsilon_m + i\eta}
      -\frac{\braket{\psi_{\mathrm{pu}}(\tau)|n}  \braket{n|\hat{J}^\alpha_{\mathrm{T},{\bf k}}|l}  \braket{l|\hat{J}^\beta_{\mathrm{T},-{\bf k}'}|m}  \braket{m|\psi_{\mathrm{pu}}(\tau)}}{\hbar \omega + \epsilon_n - \epsilon_l + i\eta}
      \right]   \nonumber \\
    &\hspace{1em}\times e^{i\omega \tau} A_{\mathrm{pr},{\bf k}'}^\beta \left( \omega + \frac{\epsilon_n - \epsilon_m}{\hbar}, 0 \right).
    \label{eq:DeltaJ_psipu}
  \end{align}
\end{widetext}
Here, $\ket{\psi_{\mathrm{pu}}(\tau)}$ is the state in the absence of the probe pulse, and is defined as
\begin{equation}
  \ket{\psi_{\mathrm{pu}}(\tau)} \coloneq \exp \left[ -i \frac{\hat{H}_0}{\hbar}(\tau-t_0) \right] \ket{\psi_0}.
  \label{eq:psi_pu}
\end{equation}
We assume that the pump pulse does not cause linear absorption, but only causes transitions from the ground state to biexciton eigenstates via two-photon absorption processes.
$\ket{\psi_{\mathrm{pu}}(\tau)}$ can be expressed as
\begin{equation}
  \ket{\psi_{\mathrm{pu}}(\tau)} = \sqrt{W_0}\exp \left( -i\frac{\epsilon_0}{\hbar}\tau \right) \ket{0} + \sqrt{W_{\mathrm{bi}}}\ket{\psi_{\mathrm{bi}}(\tau)},
  \label{eq:def_psi_bi}
\end{equation}
where the second term is the biexciton component of $\ket{\psi_{\mathrm{pu}}(\tau)}$, and $\ket{\psi_{\mathrm{bi}}(\tau)}$ is normalized: $\braket{\psi_{\mathrm{bi}}(\tau)|\psi_{\mathrm{bi}}(\tau)} = 1$.
$W_0$ $(W_{\mathrm{bi}})$ is the weight of the ground (biexciton) state.
We suppose that the spectral width of the pump pulse is so narrow that the energy of the excited biexciton can be regarded as nearly constant:
\begin{equation}
  \epsilon_m - \epsilon_0 \simeq \epsilon_{\mathrm{bi}} \ \ \mathrm{if} \braket{m|\psi_{\mathrm{bi}}(\tau)}\neq 0.
  \label{eq:eigenvalue_biexciton}
\end{equation}
Similarly, the wavevector of the biexciton is assumed to be nearly constant, which can be formulated using the spatial translation operator $\hat{T}_{{\bf R}}$,
\begin{equation}
  \hat{T}_{{\bf R}} \ket{\psi_{\mathrm{bi}}(\tau)} \simeq \exp(-2i{\bf k}_{\mathrm{pu}}\cdot {\bf R})\ket{\psi_{\mathrm{bi}}(\tau)},
  \label{eq:wavevector_biexciton}
\end{equation}
where ${\bf k}_{\mathrm{pu}} = (k_{\mathrm{pu},x}, k_{\mathrm{pu},y}, k_{\mathrm{pu},z})$ is the wavevector of the pump field in the sample.
Here, the spatial translation operator is defined as
\begin{equation}
  \hat{T}_{\bf R}\hat{O}({\bf r})\hat{T}_{\bf R}^{-1} = \hat{O}({\bf r}+{\bf R}),\ \  \hat{T}_{\bf R}\ket{0} = \ket{0},
  \label{eq:TR_def}
\end{equation}
where $\hat{O}({\bf r})$ is an arbitrary operator depending on position ${\bf r}$, and ${\bf R}$ is an arbitrary lattice vector.

By using Eqs.~(\ref{eq:DeltaJ_psipu}), (\ref{eq:def_psi_bi}), (\ref{eq:eigenvalue_biexciton}), and (\ref{eq:wavevector_biexciton}), FWM component $J_{\mathrm{T},\mathrm{FWM},{\bf k}}^\alpha$ extracted from the probe-induced transverse current $\Delta J_{\mathrm{T},{\bf k}}^\alpha$ can be obtained as
\begin{align}
  &J_{\mathrm{T},\mathrm{FWM},{\bf k}}^\alpha (\omega,\tau)
  =
  V\sqrt{W_0 W_{\mathrm{bi}}}
  \exp\left[i \left(\omega + \frac{\epsilon_0}{\hbar} \right) \tau \right] \nonumber \\
  &\times \sum_{\beta = s,p}
  \braket{\phi_{{\bf k}}^{\alpha\beta}(\omega)|\psi_{\mathrm{bi}}(\tau)} 
  \left[
    A^\beta_{\mathrm{pr},2{\bf k}_{\mathrm{pu}}-{\bf k}}
    \left(  \frac{\epsilon_{\mathrm{bi}}}{\hbar}-\omega, 0   \right)
    \right]^* ,
  \label{eq:J_final}
\end{align}
where $\bra{\phi_{{\bf k}}^{\alpha \beta}(\omega)}$ is defined as
\begin{align}
  \bra{\phi_{{\bf k}}^{\alpha \beta}(\omega)} &\coloneq \bra{0}\hat{J}^\beta_{\mathrm{T},2{\bf k}_{\mathrm{pu}}-{\bf k}}  \frac{1}{\hbar \omega - \epsilon_{\mathrm{bi}}+(\hat{H}_0-\epsilon_0)+i\eta}\hat{J}^\alpha_{\mathrm{T},{\bf k}}  \nonumber \\
  &\hspace{1em}-\bra{0} \hat{J}^\alpha_{\mathrm{T},{\bf k}}  \frac{1}{\hbar \omega - (\hat{H}_0-\epsilon_0)+i\eta} \hat{J}^\beta_{\mathrm{T},2{\bf k}_{\mathrm{pu}}-{\bf k}}.
\end{align}
For $\omega > 0$, terms with poles at $\Re \omega < 0$ do not contribute and are therefore neglected.
By substituting Eq.~(\ref{eq:J_final}) into Eq.~(\ref{eq:E_FWM}), a semiclassical representation of the FWM-electric field transmitted from the sample can be obtained:
\begin{align}
  E_{\mathrm{FWM},{\bf k}_{\parallel}}^\alpha(\omega,\tau) &= 
  t_{\mathrm{sv},{\bf k}_0}^\alpha (\omega)V\sqrt{W_0 W_{\mathrm{bi}}}e^{i \left(\omega + \frac{\epsilon_0}{\hbar} \right) \tau} \nonumber \\
  &\times \sum_{k_z}\sum_{\beta = s,p} f_{{\bf k}}(\omega)  \braket{\phi_{{\bf k}}^{\alpha\beta}(\omega)|\psi_{\mathrm{bi}}(\tau)} \nonumber \\
  &\times 
  \left[
    A^\beta_{\mathrm{pr},2{\bf k}_{\mathrm{pu}}-{\bf k}}
    \left(  \frac{\epsilon_{\mathrm{bi}}}{\hbar}-\omega, 0   \right)
    \right]^* .
  \label{eq:E_FWM_biex}
\end{align}

%% file: DMS_SecII-B.tex
\subsection{\label{SecIIB}Polarization tomography}
Hereafter, we focus on the polarization- and $\tau$-dependence of the FWM signal.
First, we derive a relation between ${\bf A}_{\mathrm{pr},{\bf k}}$ and the incident vector potential ${\bf A}_{\mathrm{in}}$ in vacuum, which can be expressed as
\begin{equation}
  A_{\mathrm{in}}^\alpha({\bf r},\omega,\tau) = \sum_{{\bf k}_{\parallel}}A_{\mathrm{in},{\bf k}_{\parallel}}^\alpha(\omega,\tau) e^{
    i {\bf k}_{\mathrm{tr}}\cdot {\bf r} 
    } .
\end{equation}
The vector potential of the probe pulse in the sample becomes
\begin{equation}
  A_{\mathrm{pr}}^\alpha({\bf r},\omega,\tau) =
  \sum_{{\bf k}_{\parallel}} t_{\mathrm{vs},{\bf k}_{\mathrm{tr}}}^\alpha A_{\mathrm{in},{\bf k}_{\parallel}}^{\alpha}(\omega,\tau) e^{i{\bf k}_{0}\cdot {\bf r}},
\end{equation}
where $t_{\mathrm{vs},{\bf k}_{\mathrm{tr}}}^\alpha$ is an amplitude transmittance of $\alpha$-polarized light with the wavevector ${\bf k}_{\mathrm{tr}}$ from vacuum to the sample. Note that the subscript ``vs" (``sv") means that the light is transmitted from vacuum to the sample (from the sample to vacuum).
Using the Fourier transform, we can obtain 
\begin{align}
  A^\alpha_{\mathrm{pr},{\bf k}}(\omega,\tau)
  &=
  \frac{1}{L_z}\int^{L_z}_{0}\mathrm{d}z\, t_{\mathrm{vs},{\bf k}_{\mathrm{tr}}}^\alpha A_{\mathrm{in},{\bf k}_{\parallel}}^{\alpha}(\omega,\tau) e^{i(k_0-k_z)z}
  \nonumber \\
  &= 
  t_{\mathrm{vs},{\bf k}_{\mathrm{tr}}}^\alpha A_{\mathrm{in},{\bf k}_{\parallel}}^{\alpha}(\omega,\tau) \left[
    \frac{e^{i(k_0-k_z)L_z}-1}{i(k_0-k_z)L_z}
    \right].
    \label{eq:Apr}
\end{align}

Next, we introduce the polarization vectors ${\bf e}_1$ and ${\bf e}_2$ (${\bf e}_i \in \mathbb{C}^3$, $|{\bf e}_i|=1$) as follows. For polarization-resolved measurements of FWM light, ${\bf e}_1$ denotes the measurement polarization, and is expanded as ${\bf e}_1 = \sum_{\alpha = s,p}e_1^\alpha {\bf e}_{{\bf k}_{\mathrm{tr}}}^\alpha$. The measured intensity can be expressed as $|{\bf e}_{1}\cdot {\bf E}_{\mathrm{FWM},{\bf k}_{\parallel}}(\omega,\tau)|^2$. 
The incident probe vector potential which generates ${\bf E}_{\mathrm{FWM},{\bf k}_{\parallel}}(\omega,\tau)$ is
${\bf A}_{\mathrm{in},2{\bf k}_{\mathrm{pu},\parallel}-{\bf k}_{\parallel}}(\epsilon_{\mathrm{bi}}/\hbar-\omega,\tau)$ where ${\bf k}_{\mathrm{pu},\parallel}\coloneq k_{\mathrm{pu},x}{\bf e}_x + k_{\mathrm{pu},y} {\bf e}_y$.
Its polarization is denoted by ${\bf e}_2$ and expanded as ${\bf e}_2 =  \sum_{\beta=s,p} e_2^\beta {\bf e}_{{\bf k}_{\mathrm{pr},\mathrm{tr}}}^\beta$ with ${\bf k}_{\mathrm{pr,tr}} \coloneq 2{\bf k}_{\mathrm{pu},\parallel} - {\bf k}_{\parallel}+ {\bf e}_z \sqrt{(\epsilon_{\mathrm{bi}}/\hbar-\omega)^2/c^2-|2{\bf k}_{\mathrm{pu},\parallel}-{\bf k}_{\parallel}|^2}$.
By using Eqs.~(\ref{eq:E_FWM_biex}) and (\ref{eq:Apr}), the polarized intensity of the FWM signal takes the form
\begin{equation}
  |{\bf e}_{1}\cdot {\bf E}_{\mathrm{FWM},{\bf k}_{\parallel}}(\omega,\tau)|^2 
  \simeq \mathcal{N} \left |
  \sum_{\alpha,\beta=s,p} (e_{1}^\alpha e_{2}^\beta)^* 
  \psi_{{\bf k}_{\parallel}}^{\alpha \beta}(\omega,\tau)
  \right |^2,
  \label{eq:polarized-intensity}
\end{equation}
where the dot ``$\cdot$'' denotes the Hermitian inner product, ${\bf a} \cdot {\bf b} \equiv {\bf a}^\dagger {\bf b}$, $\mathcal{N}$ is a polarization-independent factor, and $\psi_{{\bf k}_{\parallel}}^{\alpha \beta}(\omega,\tau)$ is defined as
\begin{equation}
 \psi_{{\bf k}_{\parallel}}^{\alpha \beta}(\omega,\tau) \coloneq  
 \frac{t_{\mathrm{sv},{\bf k}_0}^\alpha  \left(t_{\mathrm{vs},{\bf k}_{\mathrm{pr,tr}}}^\beta \right)^* \braket{\phi_{{\bf k}^*}^{\alpha\beta}(\omega)|\psi_{\mathrm{bi}}(\tau)}}{\sqrt{\sum_{\alpha'\beta'}
      \left| t_{\mathrm{sv},{\bf k}_0}^{\alpha'} \right|^2
      \left| t_{\mathrm{vs},{\bf k}_{\mathrm{pr,tr}}}^{\beta'} \right|^2
      \left |
      \braket{\phi_{{\bf k}^*}^{\alpha'\beta'}(\omega)|\psi_{\mathrm{bi}}(\tau)}
      \right |^2 }},
      \label{eq:sec2_psi_thick}
\end{equation}
where ${\bf k}^* = (k_x,k_y,k_z^*)$, and $k_z^*$ is defined as the $k_z$ that minimizes the phase mismatch $|\Delta k| = |k_z-k_0|$.
In the derivation of Eq.~(\ref{eq:polarized-intensity}), we apply the approximation that $\braket{\phi_{{\bf k}}^{\alpha\beta}(\omega)|\psi_{\mathrm{bi}}(\tau)}$ is replaced with $\braket{\phi_{{\bf k}^*}^{\alpha\beta}(\omega)|\psi_{\mathrm{bi}}(\tau)}$ in Eq.~(\ref{eq:E_FWM_biex}), which would be justified because the term with the smallest phase mismatch $|\Delta k|$ gives the dominant contribution in the sum over $k_z$ in Eq.~(\ref{eq:E_FWM_biex}).
Equation~(\ref{eq:polarized-intensity}) can be rewritten as
\begin{equation}
  |{\bf e}_1\cdot {\bf E}_{\mathrm{FWM},{\bf k}_{\parallel}}(\omega,\tau)|^2 
  \simeq \mathcal{N}
  \Tr \left[ \hat{\rho}_{\mathrm{DMS}}(\tau) \hat{\mu} \right],
  \label{eq:sec2_tomography}
\end{equation}
where $\hat{\rho}_{\mathrm{DMS}}(\tau)$ is a two-qubit density matrix defined as
\begin{align}
  &\hat{\rho}_{\mathrm{DMS}}(\tau) \equiv \ket{\psi_{\mathrm{DMS}}(\tau)} \bra{\psi_{\mathrm{DMS}}(\tau)}, \nonumber \\
  & \ket{\psi_{\mathrm{DMS}}(\tau)} \coloneq 
    \sum_{\alpha,\beta=s,p}\psi_{{\bf k}_{\parallel}}^{\alpha \beta}(\omega, \tau)  \ket{\alpha \beta},
\end{align}
with $\ket{\alpha \beta}$ an orthonormal basis for the two qubits. The operator $\hat{\mu}$ is a projection operator,
\begin{align} 
 &\hat{\mu} \equiv \ket{{\bf e}_1{\bf e}_2}\bra{{\bf e}_1{\bf e}_2}, \nonumber \\
 &\ket{{\bf e}_1{\bf e}_2} \coloneq \left( \sum_{\alpha=s,p}e_{1}^\alpha \ket{\alpha} \right) \otimes \left( \sum_{\beta=s,p}e_{2}^\beta \ket{\beta} \right).
\end{align}
The right-hand side of Eq.~(\ref{eq:sec2_tomography}) gives the measurement probability defined by $\hat{\mu}$ for the state $\hat{\rho}_{\mathrm{DMS}}(\tau)$.
By changing the polarization vectors $\{ {\bf e}_1, {\bf e}_2 \}$ and repeating the measurement of the polarized FWM intensity, we obtain a dataset $\{\Tr \left[ \hat{\rho}_{\mathrm{DMS}}(\tau) \hat{\mu} \right]\}_{\hat{\mu}}$.
After collecting sufficient experimental data, we can reconstruct $\hat{\rho}_{\mathrm{DMS}}(\tau)$ via the well-known two-qubit tomography~\cite{Tomography}.
Namely, $\psi^{\alpha \beta}_{{\bf k}_{\parallel}}(\omega,\tau)$ can be determined experimentally except for a polarization-independent phase factor.
In this way, we can obtain some information about the biexciton state $\ket{\psi_{\mathrm{bi}}(\tau)}$.

When the effects of multiple reflections are taken into account, the definition of $\psi_{{\bf k}_{\parallel}}^{\alpha \beta}(\omega,\tau)$ is changed (see Appendix~\ref{AppendixB}).

%% file: DMS_SecII-C.tex
\subsection{\label{SecIIC}Example}
To clarify the relation between $\psi_{{\bf k}_{\parallel}}^{\alpha \beta}(\omega, \tau)$ and $\ket{\psi_{\mathrm{bi}}(\tau)}$, we provide a simple example.
For simplicity, we consider only two degenerate modes of transverse excitons in an isotropic three-dimensional crystal.
The transverse current operator may be described as
\begin{equation}
  \hat{J}_{\mathrm{T},{\bf k}}^\alpha = j_{\bf k} \hat{b}_{{\bf k}\alpha} + j_{-{\bf k}}^* \hat{b}_{-{\bf k}\alpha}^{\dagger},
\end{equation}
where $\hat{b}_{{\bf k}\alpha}$ $(\hat{b}_{{\bf k}\alpha}^\dagger)$ is the annihilation (creation) operator of a transverse exciton with wavevector ${\bf k}$ and polarization $\alpha$.
These annihilation and creation operators obey bosonic commutation relations, and $\hat{b}_{{\bf k}\alpha}\ket{0} = 0$ holds.
We suppose that $\hat{b}_{{\bf k}\alpha}^\dagger \ket{0}$ is an eigenstate of $\hat{H}_0-\epsilon_0$ with eigenenergy $\hbar \Omega_{\mathrm{T},{\bf k}}$.
Hereafter, we neglect the difference between ${\bf k}^*$ and ${\bf k}_0$, and assume that the phase-matching condition ${\bf k}_0 = 2{\bf k}_{\mathrm{pu}}-{\bf k}_{\mathrm{pr}}$ is approximately satisfied, where ${\bf k}_{\mathrm{pr}} \coloneq 2{\bf k}_{\mathrm{pu},\parallel} - {\bf k}_{\parallel}+ {\bf e}_z \sqrt{[n(\epsilon_{\mathrm{bi}}/\hbar - \omega)]^2 (\epsilon_{\mathrm{bi}}/\hbar - \omega)^2/c^2-|2{\bf k}_{\mathrm{pu},\parallel}-{\bf k}_{\parallel}|^2}$.
In this setup, $\braket{\phi_{{\bf k}_0}^{\alpha\beta}(\omega)|\psi_{\mathrm{bi}}(\tau)}$ in Eq.~\eqref{eq:sec2_psi_thick} becomes
\begin{align}
  \braket{\phi_{{\bf k}_0}^{\alpha\beta}(\omega)|\psi_{\mathrm{bi}}(\tau)}
  &=\frac{j_{{\bf k}_0}j_{{\bf k}_{\mathrm{pr}}}\left[\epsilon_{\mathrm{bi}}-\hbar(\Omega_{\mathrm{T},{\bf k}_0}+\Omega_{\mathrm{T},{\bf k}_{\mathrm{pr}}}) \right]}{(\hbar \omega + \hbar \Omega_{\mathrm{T},{\bf k}_{\mathrm{pr}}} - \epsilon_{\mathrm{bi}})(\hbar \omega - \hbar \Omega_{\mathrm{T},{\bf k}_0})}  \nonumber \\
  &\hspace{1em}\times \braket{0| \hat{b}_{{\bf k}_0 \alpha} \hat{b}_{{\bf k}_{\mathrm{pr}} \beta} |\psi_{\mathrm{bi}}(\tau)}.
\end{align}
The factor $\braket{0| \hat{b}_{{\bf k}_0 \alpha} \hat{b}_{{\bf k}_{\mathrm{pr}} \beta} |\psi_{\mathrm{bi}}(\tau)}$ is one of the expansion coefficients of the biexciton wavefunction  $\ket{\psi_{\mathrm{bi}}(\tau)}$ in terms of a superposition of products of two exciton wavefunctions. 
By defining $\hat{P}_{{\bf k}_0,{\bf k}_{\mathrm{pr}}}$ as the projection operator onto the subspace $\spanspace \{ \hat{b}_{{\bf k}_0\alpha}^\dagger \hat{b}_{{\bf k}_{\mathrm{pr}}\beta}^\dagger \ket{0} \}_{\alpha,\beta=s,p}$, 
the projected state $\hat{P}_{{\bf k}_0,{\bf k}_{\mathrm{pr}}} \ket{\psi_{\mathrm{bi}}(\tau)}$ can be expressed as
\begin{equation}
  \hat{P}_{{\bf k}_0,{\bf k}_{\mathrm{pr}}} \ket{\psi_{\mathrm{bi}}(\tau)} \propto \sum_{\alpha \beta}
  \frac{\psi_{{\bf k}_{\parallel}}^{\alpha \beta}(\tau)}{t_{\mathrm{sv},{\bf k}_0}^\alpha \left(t_{\mathrm{vs},{\bf k}_{\mathrm{pr,tr}}}^\beta \right)^*}
  \hat{b}_{{\bf k}_0 \alpha}^\dagger \hat{b}_{{\bf k}_{\mathrm{pr}} \beta}^\dagger \ket{0}.
\end{equation}
After normalization, we obtain the biexciton state projected onto the exciton pair with wavevectors ${\bf k}_0$ and ${\bf k}_{\mathrm{pr}}$, where ${\bf k}_0$ (${\bf k}_{\mathrm{pr}}$) is the wavevector of the FWM (probe) light in the sample. Here, $t_{\mathrm{sv},{\bf k}_0}^\alpha$ and $t_{\mathrm{vs},{\bf k}_{\mathrm{pr,tr}}}^\beta$ can be calculated with the refractive index of the sample, and $\psi_{{\bf k}_{\parallel}}^{\alpha \beta}(\tau)$ can be obtained experimentally using quantum tomography, as mentioned above.

Here, we mention two limitations of DMS.
First, the information obtained by DMS is limited to the transverse exciton pair in the biexciton, i.e., the pair of one-photon-allowed excitons, although the biexciton includes components that involve one-photon-forbidden excitons.
For example, the $\Gamma_1$-biexciton state in CuCl is a superposition of a $\Gamma_2 \otimes \Gamma_2$-exciton-pair and a $\Gamma_5 \otimes \Gamma_5$-exciton-pair, where the $\Gamma_2$ exciton is optically forbidden~\cite{ExcitonicProcesses}.
In addition, the $\Gamma_5$-exciton consists of two transverse exciton modes and one longitudinal exciton mode, the latter being one-photon-forbidden. Thus, in this example, only the transverse components of the $\Gamma_5$ exciton contribute to the DMS signal.

Second, ${\bf k}_0$ and ${\bf k}_{\mathrm{pr}}$ are limited by the dispersion relation of light and the phase-matching condition of FWM.
For a CuCl crystal, the wavevectors are limited to $|{\bf k}_0|, |{\bf k}_{\mathrm{pr}}| \leq 1.3\times 10^6$ cm$^{-1}$ due to the phase-matching condition of FWM calculated using the Lorentz model with spatial dispersion (the parameters used here are those obtained experimentally in Refs.~\cite{Param1,Param2}).
On the other hand, the biexciton radius $a_{\mathrm{bi}}$ is estimated at 1.5 nm~\cite{CuCl_biradius};  therefore, $a_{\mathrm{bi}}^{-1} = 6.7\times 10^6$ cm$^{-1} > |{\bf k}_0|, |{\bf k}_{\mathrm{pr}}|$.
This relation suggests that it is difficult to extract information about the spatial extent of the biexciton from experiments using FWM such as DMS.

%% file: DMS_SecIII.tex
\section{\label{SecIII}Formulations for thin samples}
In this section, we formulate DMS for samples whose thickness $L_z$ is sufficiently small compared with the wavelength of the light pulses ($L_z \ll \lambda$).
The sample is assumed to be an uncharged insulator, and to occupy the region $-L_x/2<x<L_x/2$, $-L_y/2<y<L_y/2$, and $0<z<L_z$.
For simplicity, $L_x$ and $L_y$ are assumed to be infinitely large.
We assume the probe pulse is incident from $z<0$, and the transmitted FWM light is detected at $z \gg \lambda$. By taking the thin-film limit $L_z \to 0$, 
the current density is modeled as 
\begin{equation}
  {\bf J}({\bf r},\omega) = {\bf j}({\bf r}_{\parallel},\omega)\delta(z) = \sum_{{\bf k}_{\parallel}} {\bf j}_{{\bf k}_{\parallel}}(\omega)e^{i{\bf k}_{\parallel}\cdot {\bf r}_{\parallel}}\delta(z),
  \label{eq:J_thin}
\end{equation}
where ${\bf j} \perp \mathbf{e}_z$, ${\bf r}_{\parallel} = x{\bf e}_x + y{\bf e}_y$, and $\delta(z)$ is the Dirac delta function.
From Maxwell's equations, the electric field of the light generated by the current in the sample, $\Delta {\bf E}_{\mathrm{T}}$, takes the form:
\begin{align}
  &\Delta {\bf E}_{\mathrm{T}}({\bf r},\omega) =
  \sum_{{\bf k}_{\parallel}} \Delta {\bf E}_{\mathrm{T},{\bf k}_{\parallel}}(z,\omega) 
  e^{i{\bf k}_{\parallel}\cdot {\bf r}}, \nonumber \\
  &\Delta {\bf E}_{\mathrm{T},{\bf k}_{\parallel}}(z,\omega) \simeq  
  \frac{-\mu_0\omega}{2k_{\mathrm{tr}}}
  {\bf j}_{\mathrm{T},{\bf k}_{\mathrm{tr}}}(\omega) e^{ik_{\mathrm{tr}}z} \ (z \gg 1/|{\bf k}_{\parallel}|), \\
  &\Delta {\bf E}_{\mathrm{T},{\bf k}_{\parallel} = {\bf 0}} 
  (z,\omega) = 
  \frac{-\mu_0\omega}{2k_{\mathrm{tr}}}
  {\bf j}_{\mathrm{T},{\bf k}_{\mathrm{tr}}}(\omega) e^{ik_{\mathrm{tr}}z} \ (z>0).
   \label{eq:E_thin}
\end{align}
Here, ${\bf j}_{\mathrm{T},{\bf k}}(\omega)$ is defined as
\begin{equation}
  {\bf j}_{\mathrm{T},{\bf k}}(\omega) ={\bf j}_{{\bf k}_{\parallel}}(\omega) - \left[{\bf e}_{{\bf k}}\cdot {\bf j}_{{\bf k}_{\parallel}}(\omega)\right]{\bf e}_{{\bf k}}.
  \label{eq:j_T_def}
\end{equation}
The derivation of Eq.~(\ref{eq:E_thin}) is found in Appendix~\ref{AppendixC}.

The nonlinear current density produced via the FWM process, ${\bf J}_{\mathrm{FWM}}({\bf r},\omega)$, is described as ${\bf J}_{\mathrm{FWM}}({\bf r},\omega) = {\bf j}_{\mathrm{FWM}}({\bf r}_{\parallel},\omega)\delta(z)$. 
The emitted FWM light, ${\bf E}_{\mathrm{FWM}}$, at $z \gg \lambda$ is given by
\begin{eqnarray}
  &&{\bf E}_{\mathrm{FWM}}({\bf r},\omega) =
  \sum_{{\bf k}_{\parallel}}{\bf E}_{\mathrm{FWM},{\bf k}_{\parallel}}(\omega) 
  e^{i{\bf k}_{\mathrm{tr}}\cdot {\bf r}}, \nonumber \\
  &&{\bf E}_{\mathrm{FWM},{\bf k}_{\parallel}}(\omega) =  
  \frac{-\mu_0\omega}{2k_{\mathrm{tr}}}
  {\bf j}_{\mathrm{T,FWM},{\bf k}_{\mathrm{tr}}}(\omega).
   \label{eq:E_FWM_thin}
\end{eqnarray}

To discuss the relation between the FWM signal and the biexciton wavefunctions, we derive an analytical representation of ${\bf j}_{\mathrm{T},\mathrm{FWM},{\bf k}}$ within the perturbative approach. The pump pulse is applied to the sample at time $t=0$, and we focus on the subsequent dynamics after the excitation.
In this time interval, we consider the two-dimensional semiclassical light-matter Hamiltonian,
\begin{align}
  \hat{H}(t,\tau) &= \hat{H}_0 - \int \mathrm{d}x \int \mathrm{d}y\, \hat{{\bf j}}(x,y) \cdot {\bf A}_{\mathrm{pr}}(x,y,z=0,t,\tau) \nonumber  \\
  &= \hat{H}_0 - S\sum_{{\bf k}_{\parallel}} \hat{{\bf j}}_{{-\bf k}_{\parallel}}\cdot {\bf A}_{\mathrm{pr},{\bf k}_{\parallel}}(z=0,t,\tau),
  \label{eq:H_thin}
\end{align}
where $\hat{{\bf j}}$ is the current density operator,  ${\bf A}_{\mathrm{pr}}$ is the vector potential of the probe pulse, ${\bf A}_{\mathrm{pr},{\bf k}_{\parallel}}$ is the Fourier transform of ${\bf A}_{\mathrm{pr}}$ in terms of $x$ and $y$, and $S = L_xL_y$.
Periodic boundary conditions are imposed in the $x$ and $y$ directions.
Equation~(\ref{eq:H_thin}) can be transformed as
\begin{align}
  \hat{H}(t,\tau)  = \hat{H}_0 - S\sum_{{\bf k}_{\parallel}} \int^{\infty}_{-\infty}\mathrm{d}k_z \ \hat{{\bf j}}_{\mathrm{T},-{\bf k}} \cdot {\bf A}_{\mathrm{pr},{\bf k}}(t,\tau),
  \label{eq:H_thin2}
\end{align}
where ${\bf k} = {\bf k}_{\parallel} + k_z {\bf e}_z$, and ${\bf A}_{\mathrm{pr},{\bf k}}$ is the Fourier transform of ${\bf A}_{\mathrm{pr},{\bf k}_{\parallel}}(z,t,\tau)$ in terms of $z$: ${\bf A}_{\mathrm{pr},{\bf k}} = (1/2\pi)\int^\infty_{-\infty}\mathrm{d}z\, {\bf A}_{\mathrm{pr},{\bf k}_{\parallel}}(z)e^{-ik_z z} $ [that is slightly different from Eq.~(\ref{eq:FT-def})].
In the derivation of Eq.~(\ref{eq:H_thin2}), we use ${\bf k}\cdot{\bf A}_{\mathrm{pr},{\bf k}}(t,\tau)=0$,
which follows from the Coulomb gauge condition.
The transverse FWM current ${\bf j}_{\mathrm{T},\mathrm{FWM},{\bf k}_{\mathrm{tr}}}$ can be derived in the same way as in Sec.~\ref{SecIIA}.
By applying time-dependent perturbation theory and using Eqs.~(\ref{eq:Apr_condition}), (\ref{eq:psi_pu}), (\ref{eq:def_psi_bi}), and (\ref{eq:eigenvalue_biexciton}), we can obtain 
\begin{widetext}
  \begin{align}
    j_{\mathrm{T},\mathrm{FWM},{\bf k}_{\mathrm{tr}}}^\alpha (\omega,\tau) =&
    \sum_{\beta = s,p}\sum_{{\bf k}'_{\parallel}} \int^{\infty}_{-\infty}\mathrm{d}k'_z 
    \left[
      \bra{0}\hat{j}^\beta_{\mathrm{T},-{\bf k}'}
      \frac{1}{\hbar \omega - \epsilon_{\mathrm{bi}}+(\hat{H}_0-\epsilon_0)+i\eta} 
      \hat{j}^\alpha_{\mathrm{T},{\bf k}_{\mathrm{tr}}}\ket{\psi_{\mathrm{bi}}(\tau)}\right.
      \nonumber \\
      &\left.-\bra{0} \hat{j}^\alpha_{\mathrm{T},{\bf k}_{\mathrm{tr}}}  
      \frac{1}{\hbar \omega - (\hat{H}_0-\epsilon_0)+i\eta} \hat{j}^\beta_{\mathrm{T},-{\bf k}'} \ket{\psi_{\mathrm{bi}}(\tau)} \right] 
      S\sqrt{W_0 W_{\mathrm{bi}}} e^{i \left(\omega + \frac{\epsilon_0}{\hbar} \right) \tau} 
    \left[	
      A_{\mathrm{pr},-{\bf k}'}^\beta \left( \frac{\epsilon_{\mathrm{bi}}}{\hbar}-\omega, 0 \right)
      \right]^* \ \ \   (\omega > 0).
        \label{eq:DeltaJ_psipu_thin}
\end{align}
\end{widetext}
Equation (\ref{eq:wavevector_biexciton}) in Sec.~\ref{SecIIA} is replaced by
\begin{equation}
  \hat{T}_{{\bf R}_{\parallel}} \ket{\psi_{\mathrm{bi}}(\tau)} \simeq \exp(-2i{\bf k}_{\mathrm{pu},\parallel}\cdot {\bf R}_{\parallel})\ket{\psi_{\mathrm{bi}}(\tau)},
  \label{eq:k_pu_thin}
\end{equation}
where ${\bf k}_{\mathrm{pu},\parallel} = k_{\mathrm{pu},x} {\bf e}_x + k_{\mathrm{pu},y}{\bf e}_y $
is the wavevector of the pump field projected onto the $xy$ plane, and ${\bf R}_{\parallel}$ is an arbitrary two-dimensional lattice vector.
From Eq.~\eqref{eq:k_pu_thin}, it is deduced that nonzero terms in Eq.~\eqref{eq:DeltaJ_psipu_thin} are limited to $-{\bf k}'_{\parallel} = 2{\bf k}_{\mathrm{pu},\parallel} - {\bf k}_{\parallel}$.
In addition, if $A_{\mathrm{pr},-{\bf k}'}^\beta ( \epsilon_{\mathrm{bi}}/\hbar -\omega, 0)$ is nonzero, $-k_z' = \sqrt{(\epsilon_{\mathrm{bi}}/\hbar - \omega)^2/c^2-|{\bf k}'_{\parallel}|^2}$ holds because of the dispersion relation of light.
In this way, the ${\bf k}'_{\parallel}$ sum and the $k'_z$ integral in Eq.~\eqref{eq:DeltaJ_psipu_thin} can be carried out, and it follows that 
\begin{align}
  &j_{\mathrm{T},\mathrm{FWM},{\bf k}_{\mathrm{tr}}}^\alpha (\omega,\tau)
  =
  S\sqrt{W_0 W_{\mathrm{bi}}}
  \exp\left[i \left(\omega + \frac{\epsilon_0}{\hbar} \right) \tau \right] \nonumber \\
  & \times \sum_{\beta = s,p}
  \braket{\phi_{{\bf k}_{\mathrm{tr}}}^{\alpha\beta}(\omega)} |\psi_{\mathrm{bi}}(\tau) 
  \left[
    A^\beta_{\mathrm{pr},2{\bf k}_{\mathrm{pu},\parallel}-{\bf k}_{\parallel}}
    \left(z=0,  \frac{\epsilon_{\mathrm{bi}}}{\hbar}-\omega, 0   \right)
    \right]^* .
  \label{eq:J_final_thin}
\end{align}
Here, $\bra{\phi_{{\bf k}_{\mathrm{tr}}}^{\alpha\beta}(\omega)}$ is redefined as 
\begin{align}
  \bra{\phi_{{\bf k}_{\mathrm{tr}}}^{\alpha\beta}(\omega)} \coloneq &\bra{0}\hat{j}^\beta_{\mathrm{T},{\bf k}_{\mathrm{pr,tr}}}  \frac{1}{\hbar \omega - \epsilon_{\mathrm{bi}}+(\hat{H}_0-\epsilon_0)+i\eta}\hat{j}^\alpha_{\mathrm{T},{\bf k}_{\mathrm{tr}}}  \nonumber \\
  &-\bra{0} \hat{j}^\alpha_{\mathrm{T},{\bf k}_{\mathrm{tr}}}  \frac{1}{\hbar \omega - (\hat{H}_0-\epsilon_0)+i\eta} \hat{j}^\beta_{\mathrm{T},{\bf k}_{\mathrm{pr,tr}}}.
  \label{eq:phi_def_thin}
\end{align}
By substituting Eq.~(\ref{eq:J_final_thin}) into Eq.~(\ref{eq:E_FWM_thin}), a semiclassical representation of the FWM electric field can be obtained:
\begin{align}
  &E_{\mathrm{FWM},{\bf k}_{\parallel}}^\alpha (\omega,\tau) = 
  \frac{-\mu_0 \omega}{2k_{\mathrm{tr}}} 
  S\sqrt{W_0 W_{\mathrm{bi}}}
  \exp\left[i \left(\omega + \frac{\epsilon_0}{\hbar} \right) \tau \right] \nonumber \\
  & \times \sum_{\beta = s,p}
  \braket{\phi_{{\bf k}_{\mathrm{tr}}}^{\alpha\beta}(\omega)|\psi_{\mathrm{bi}}(\tau)}
   \left[
    A^\beta_{\mathrm{pr},2{\bf k}_{\mathrm{pu},\parallel}-{\bf k}_{\parallel}}
    \left(z=0,  \frac{\epsilon_{\mathrm{bi}}}{\hbar}-\omega, 0   \right)
    \right]^* .
  \label{eq:E_FWM_final_thin}
\end{align}

Next, we introduce the polarization vectors ${\bf e}_1$ and ${\bf e}_2$ (${\bf e}_i \in \mathbb{C}^3$, $|{\bf e}_i|=1$) 
for the thin-sample case, analogously to Sec.~\ref{SecIIB}. Here, ${\bf e}_1$ denotes the measured polarization of ${\bf E}_{\mathrm{FWM},{\bf k}_{\parallel}}(\omega,\tau)$, while ${\bf e}_2$ corresponds to the polarization of ${\bf A}_{\mathrm{pr},2{\bf k}_{\mathrm{pu},\parallel}-{\bf k}_{\parallel}}(z=0,  \epsilon_{\mathrm{bi}}/\hbar-\omega, \tau)$.
By using Eq.~(\ref{eq:E_FWM_final_thin}), the polarized intensity of the FWM signal takes the form
\begin{equation}
  |{\bf e}_{1}\cdot {\bf E}_{\mathrm{FWM},{\bf k}_{\parallel}}|^2 
  = \mathcal{N} \left |
  \sum_{\alpha,\beta=s,p} (e_{1}^\alpha e_{2}^\beta)^* 
  \psi_{{\bf k}_{\parallel}}^{\alpha \beta}(\omega,\tau)
  \right |^2,
  \label{eq:sec3_DMS_thin}
\end{equation}
where $\mathcal{N}$ is a polarization-independent factor, and $\psi_{{\bf k}_{\parallel}}^{\alpha \beta}(\omega, \tau)$ is defined as
\begin{equation}
  \psi^{\alpha \beta}_{{\bf k}_{\parallel}}(\omega,\tau) \coloneq \frac{\braket{\phi_{{\bf k}_{\mathrm{tr}}}^{\alpha\beta}(\omega)|\psi_{\mathrm{bi}}(\tau)}}{\sqrt{\sum_{\alpha'\beta'}
      \left |
      \braket{\phi_{{\bf k}_{\mathrm{tr}}}^{\alpha'\beta'}(\omega)|\psi_{\mathrm{bi}}(\tau)}
      \right |^2 }}.
      \label{eq:sec3_psi_thin}
\end{equation}
Equation~(\ref{eq:sec3_DMS_thin}) can be rewritten as
\begin{equation}
   |{\bf e}_{1}\cdot {\bf E}_{\mathrm{FWM},{\bf k}_{\parallel}}|^2
   =\mathcal{N}
  \Tr \left[ \hat{\rho}_{\mathrm{DMS}}(\tau) \hat{\mu} \right],
  \label{eq:sec3_tomography}
\end{equation}
where $\hat{\rho}_{\mathrm{DMS}}(\tau)$ is a two-qubit density matrix defined as
\begin{align}
  \hat{\rho}_{\mathrm{DMS}}(\tau) &\equiv \ket{\psi_{\mathrm{DMS}}(\tau)} \bra{\psi_{\mathrm{DMS}}(\tau)}, \nonumber \\
   \ket{\psi_{\mathrm{DMS}}(\tau)} &\coloneq 
    \sum_{\alpha,\beta=s,p}\psi_{{\bf k}_{\parallel}}^{\alpha \beta}(\omega, \tau)  \ket{\alpha \beta},
\end{align}
with an orthonormal basis for the two qubits $\ket{\alpha \beta}$, and $\hat{\mu}$ is defined as
\begin{eqnarray} 
 &&\hat{\mu} \equiv \ket{{\bf e}_1{\bf e}_2}\bra{{\bf e}_1{\bf e}_2}, \nonumber \\
 &&\ket{{\bf e}_1{\bf e}_2} \coloneq \left( \sum_{\alpha=s,p}e_{1}^\alpha \ket{\alpha} \right) \otimes \left( \sum_{\beta=s,p}e_{2}^{\beta}\ket{\beta} \right).
\end{eqnarray}
Experimentally, $\hat{\rho}_{\mathrm{DMS}}(\tau)$ can be obtained in the same way as in Sec.~\ref{SecIIB}.
Therefore, $\psi_{{\bf k}_{\parallel}}^{\alpha \beta}(\omega, \tau)$ can be obtained experimentally up to a global phase factor.

%% file: DMS_SecIV-A.tex
\section{\label{SecIV}Numerical simulations of DMS for a two-dimensional electron-hole system}

\subsection{\label{SecIVA}Model}
\begin{figure}[tb]
\includegraphics[width=6cm]{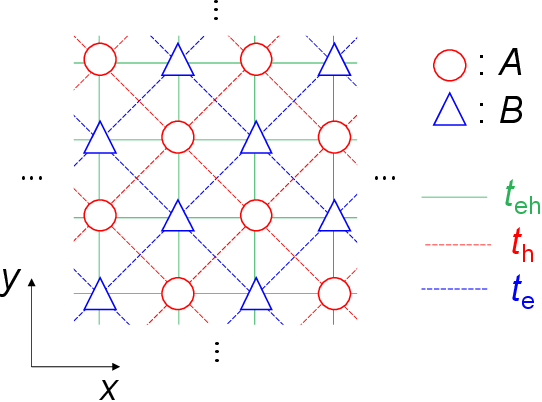}
\caption{\label{fig:model} Two-dimensional square lattice with alternating $A$ sites and $B$ sites. Holes (electrons) can occupy $A$ ($B$) sites. The hopping between nearest $A$-$B$ sites ($t_{\mathrm{eh}}$), nearest $A$-$A$ sites ($t_{\mathrm{h}}$), and nearest $B$-$B$ sites ($t_{\mathrm{e}}$) are shown by solid green line, red dashed line, and blue dashed line, respectively.}
\end{figure}

In this section, we perform numerical simulations of DMS for a two-dimensional electron-hole system described by an extended ionic Hubbard model.
We consider a square lattice in the $xy$ plane ($z=0$) with alternating $A$ and $B$ sites, occupied by holes and electrons, respectively, as shown in Fig.~\ref{fig:model}.
The number of sites $N$ is $N = 2M\times 2M$, and periodic boundary conditions are imposed in the $x$ and $y$ directions.
The distance between the nearest-neighbor sites is denoted by $a$.
A site at position $a{\bf r} = a(x{\bf e}_x + y{\bf e}_y)$ $(x,y\in \mathbb{Z})$ belongs to the $A$ ($B$) sublattice if $x+y$ is even (odd).
The model includes $s$-orbital electrons and $p$-orbital holes on this lattice.
The Hamiltonian is given by
\begin{align}
  &\hat{H}_{\mathrm{mat}} = \hat{H}_{\Delta} + \hat{H}_U + \hat{H}_{V} + \hat{H}_{t_{\mathrm{eh}}} + \hat{H}_{t_\mathrm{e}} + \hat{H}_{t_\mathrm{h}},  \\
  &\hat{H}_{\Delta} = \Delta \sum_{{\bf r}_\mathrm{e}} \hat{n}_{{\bf r}_\mathrm{e}}, \\
  &\hat{H}_U =  U\sum_{{\bf r}_\mathrm{e}} \hat{n}_{{\bf r}_\mathrm{e},\uparrow}\hat{n}_{{\bf r}_\mathrm{e},\downarrow}
  + U\sum_{{\bf r}_\mathrm{h}} \frac{\hat{n}_{{\bf r}_\mathrm{h}}^{\mathrm{h}}(\hat{n}_{{\bf r}_\mathrm{h}}^{\mathrm{h}}-1)}{2}  , \\
  &\hat{H}_V =  \frac{1}{2}\sum_{{\bf r}_\mathrm{e}\neq {\bf r}_\mathrm{e}'}\frac{V}{|{\bf r}_\mathrm{e} - {\bf r}_\mathrm{e}'|}\hat{n}_{{\bf r}_\mathrm{e}}\hat{n}_{{\bf r}_\mathrm{e'}}
  + \frac{1}{2}\sum_{{\bf r}_\mathrm{h}\neq {\bf r}_\mathrm{h}'}\frac{V}{|{\bf r}_\mathrm{h} - {\bf r}_\mathrm{h}'|}
  \hat{n}_{{\bf r}_\mathrm{h}}^{\mathrm{h}} \hat{n}_{{\bf r}_\mathrm{h}'}^{\mathrm{h}} \nonumber \\
  &\hspace{3em}-\sum_{{\bf r}_\mathrm{e}{\bf r}_\mathrm{h}}\frac{V}{|{\bf r}_\mathrm{e} - {\bf r}_\mathrm{h}|}\hat{n}_{{\bf r}_\mathrm{e}} \hat{n}_{{\bf r}_\mathrm{h}}^{\mathrm{h}},
  \\
  &\hat{H}_{t_{\mathrm{eh}}} =  t_{\mathrm{eh}}\sum_{{\bf r}_\mathrm{h}\sigma}\sum_{u=\pm 1}(-u) \left[
    \hat{c}^\dagger_{{\bf r}_\mathrm{h}+u{\bf e}_x,\sigma}\hat{h}^\dagger_{{\bf r}_\mathrm{h},-\sigma,p_x} 
    \right. \nonumber \\ 
    &\hspace{8em}\left.
    +\hat{c}^\dagger_{{\bf r}_\mathrm{h}+u{\bf e}_y,\sigma}\hat{h}^\dagger_{{\bf r}_\mathrm{h},-\sigma,p_y} + \mathrm{h.c.}
    \right], \\
  &\hat{H}_{t_\mathrm{e}}  = -t_\mathrm{e}\sum_{{\bf r}_\mathrm{e}\sigma}\sum_{u,v=\pm 1} \hat{c}^\dagger_{{\bf r}_\mathrm{e}+u{\bf e}_x + v{\bf e}_y,\sigma}\hat{c}_{{\bf r}_\mathrm{e},\sigma} , \\
  & \hat{H}_{t_\mathrm{h}}= -t_\mathrm{h}\sum_{{\bf r}_\mathrm{h}\sigma}\sum_{u=\pm 1} 
  \left[\hat{h}^\dagger_{{\bf r}_\mathrm{h}+u({\bf e}_x + {\bf e}_y),\sigma,p_{D}}\hat{h}_{{\bf r}_\mathrm{h},\sigma,p_{D}} \right. \nonumber \\
  &\hspace{8em}\left. + \hat{h}^\dagger_{{\bf r}_\mathrm{h}+u({\bf e}_x - {\bf e}_y),\sigma,p_{\bar{D}}}\hat{h}_{{\bf r}_\mathrm{h},\sigma,p_{\bar{D}}}  \right].
\end{align}
Here, $\hat{c}_{{\bf r}_\mathrm{e},\sigma}$ is the annihilation operator of an electron at site ${\bf r}_{\mathrm{e}}$ with spin $\sigma$, and $\hat{h}_{{\bf r}_\mathrm{h},\sigma,p_{\alpha}}$ is the annihilation operator of a hole at site ${\bf r}_{\mathrm{h}}$ with spin $\sigma$ and orbital $p_{\alpha}$.
The number operators of holes and electrons are defined as $\hat{n}_{{\bf r}_\mathrm{e}} = \sum_{\sigma}\hat{c}^\dagger_{{\bf r}_\mathrm{e}\sigma} \hat{c}_{{\bf r}_\mathrm{e}\sigma}$ and $\hat{n}_{{\bf r}_\mathrm{h}}^{\mathrm{h}} = \sum_{\sigma p} \hat{h}^\dagger_{{\bf r}_\mathrm{h}\sigma p} \hat{h}_{{\bf r}_\mathrm{h}\sigma p}$, respectively.
We introduce $\Delta$ as the on-site potential at the $B$ sites, $U$ as the on-site Coulomb repulsion, and $V$ as the intersite Coulomb interaction parameter.
Under the periodic boundary conditions in the $x$ and $y$ directions, the distances $|{\bf r}_{\mathrm{e}} - {\bf r}'_{\mathrm{e}}|$, $|{\bf r}_{\mathrm{h}} - {\bf r}'_{\mathrm{h}}|$, and $|{\bf r}_{\mathrm{e}} - {\bf r}_{\mathrm{h}}|$ appearing in $\hat{H}_V$ are evaluated using the shortest separation among all periodic replicas.
The parameters $t_{\mathrm{eh}}$, $t_{\mathrm{e}}$, and $t_{\mathrm{h}}$ denote the transfer integrals between the nearest-neighbor $A$ and $B$ sites, between the nearest $B$ sites, and between the nearest $A$ sites, the latter involving $\sigma$-bonding (the $\pi$-bonding terms are ignored).
The signs of these hopping terms are determined by the symmetry of the orbitals. We introduce
$p_{D}$- and $p_{\bar{D}}$-orbitals, which are defined as $(p_x+p_y)/\sqrt{2}$ and $(p_x-p_y)/\sqrt{2}$, respectively, to express $\sigma$-bonding hopping terms in $\hat{H}_{t_{\mathrm{h}}}$.
The annihilation operators of $p_{D}$- and $p_{\bar{D}}$-holes are defined as
\begin{align}
  &\hat{h}_{{\bf r}_\mathrm{h},\sigma,p_{D}}  = \frac{\hat{h}_{{\bf r}_\mathrm{h},\sigma,p_{x}} + \hat{h}_{{\bf r}_\mathrm{h},\sigma,p_{y}}}{\sqrt{2}},   \\
  &\hat{h}_{{\bf r}_\mathrm{h},\sigma,p_{\bar{D}}}  = \frac{\hat{h}_{{\bf r}_\mathrm{h},\sigma,p_{x}} - \hat{h}_{{\bf r}_\mathrm{h},\sigma,p_{y}}}{\sqrt{2}}.
\end{align}

In the simulation of DMS in this section, the light pulses are incident normally on the $xy$ plane.
The wavevectors of the light pulses projected onto the $xy$ plane are set to ${\bf k}_{\parallel} = {\bf 0}$ (the effects of finite spot size are ignored), which means that the vector potential ${\bf A}(x,y,z=0,t)$ in this two-dimensional system is independent of position: ${\bf A}(x,y,z=0,t) = {\bf A}(t)$.
The light-matter interaction is introduced using the Peierls substitution, which modifies hopping terms.
For example, $-t_{\mathrm{eh}}\hat{c}^\dagger_{{\bf r}_\mathrm{h}+u{\bf e}_x,\sigma}\hat{h}^\dagger_{{\bf r}_\mathrm{h},-\sigma,p_x}$ is replaced with $-t_{\mathrm{eh}}e^{-ieauA^x(t)/\hbar}\hat{c}^\dagger_{{\bf r}_\mathrm{h}+u{\bf e}_x,\sigma}\hat{h}^\dagger_{{\bf r}_\mathrm{h},-\sigma,p_x}$, where $A^\alpha(t)$ ($\alpha = x,y$) is the $\alpha$ component of ${\bf A}(t)$, and $e$ is the elementary charge ($e>0$). 
A Taylor expansion with respect to ${\bf A}$ up to the first order under the condition $eaA^{\alpha}/\hbar \ll 1$ reduces the light-matter interaction terms to $-S\hat{{\bf j}}\cdot {\bf A}(t)$, where $\hat{{\bf j}}$ is the current density operator with $\hat{{\bf j}} = \hat{{\bf j}}_{{\bf k}_{\parallel} = {\bf 0}} =  \hat{{\bf j}}_{\mathrm{T},{\bf k}=k_z {\bf e}_z}$, and $S = Na^2$ denotes the area of the system.
The operator $\hat{{\bf j}}$ consists of three parts: $\hat{{\bf j}} = \hat{{\bf j}}_+ + \hat{{\bf j}}_- + \hat{{\bf j}}_0$, where $\hat{{\bf j}}_+$ $(\hat{{\bf j}}_-)$ increases (decreases) the number of electron-hole pairs, while $\hat{{\bf j}}_0$ conserves the number of electron-hole pairs.
$\hat{{\bf j}}_+$ and $\hat{{\bf j}}_-$ originate from the Peierls substitution of $\hat{H}_{t_{\mathrm{eh}}}$, and the explicit forms are given by
\begin{align}
  &\hat{{\bf j}}_{+}
  = \sum_{\sigma = \uparrow,\downarrow} \hat{{\bf j}}_{+,\sigma}\nonumber \\
  &=  -\frac{ieat_{\mathrm{eh}}}{S\hbar} \sum_{\bm{r}_{\mathrm{h}}\sigma}\sum_{u=\pm 1}
  \begin{bmatrix}
    \hat{c}_{\bm{r}_{\mathrm{h}}+u{\bf e}_x,-\sigma}^\dagger \hat{h}_{\bm{r}_{\mathrm{h}},\sigma,p_x}^\dagger \\
    \hat{c}_{\bm{r}_{\mathrm{h}}+u{\bf e}_y,-\sigma}^\dagger \hat{h}_{\bm{r}_{\mathrm{h}},\sigma,p_y}^\dagger
  \end{bmatrix},  \\
  &\hat{{\bf j}}_{-} = \sum_{\sigma = \uparrow,\downarrow} \hat{{\bf j}}_{-,\sigma} = \hat{\bf j}_+^\dagger .
\end{align}
$\hat{{\bf j}}_0$ comes from the Peierls substitution of $\hat{H}_{t_{\mathrm{h}}}$ and $\hat{H}_{t_{\mathrm{e}}}$ and takes the form
\begin{align}
  \hat{{\bf j}}_0
  = &-\frac{ieat_{\mathrm{e}}}{S\hbar} \sum_{\bm{r}_{\mathrm{e}} \sigma}\sum_{u,v=\pm 1}
  \begin{bmatrix}
    u \hat{c}_{\bm{r}_{\mathrm{e}}+u{\bf e}_x + v{\bf e}_y,\sigma}^\dagger \hat{c}_{\bm{r}_{\mathrm{e}},\sigma} \\
    v \hat{c}_{\bm{r}_{\mathrm{e}}+u{\bf e}_x + v{\bf e}_y,\sigma}^\dagger \hat{c}_{\bm{r}_{\mathrm{e}},\sigma}
  \end{bmatrix}  \nonumber \\
  &+\frac{ieat_{\mathrm{h}}}{S\hbar} \sum_{\bm{r}_{\mathrm{h}} \sigma}\sum_{u = \pm 1}
  \begin{bmatrix}
    u \hat{h}_{\bm{r}_{\mathrm{h}}+u({\bf e}_x + {\bf e}_y),\sigma,p_D}^\dagger \hat{h}_{\bm{r}_{\mathrm{h}},\sigma,p_D} \\
    u \hat{h}_{\bm{r}_{\mathrm{h}}+u({\bf e}_x + {\bf e}_y),\sigma,p_D}^\dagger \hat{h}_{\bm{r}_{\mathrm{h}},\sigma,p_D}
  \end{bmatrix} \nonumber \\
  &+\frac{ieat_{\mathrm{h}}}{S\hbar} \sum_{\bm{r}_{\mathrm{h}} \sigma}\sum_{u = \pm 1}
  \begin{bmatrix}
    u \hat{h}_{\bm{r}_{\mathrm{h}}+u({\bf e}_x - {\bf e}_y),\sigma,p_{\bar{D}}}^\dagger \hat{h}_{\bm{r}_{\mathrm{h}},\sigma,p_{\bar{D}}} \\
    - u \hat{h}_{\bm{r}_{\mathrm{h}}+u({\bf e}_x - {\bf e}_y),\sigma,p_{\bar{D}}}^\dagger \hat{h}_{\bm{r}_{\mathrm{h}},\sigma,p_{\bar{D}}}
  \end{bmatrix}.
\end{align}

For the numerical calculations in this section, we use exact diagonalization techniques.
One drawback is that the computational cost for calculating the eigenstates of $\hat{H}_{\mathrm{mat}}$ increases exponentially with system size $N$ because $\hat{H}_{\mathrm{mat}}$ does not conserve the number of electron-hole pairs due to $\hat{H}_{t_{\mathrm{eh}}}$ corresponding to pair creation and annihilation.
Thus, the computable system size is strongly limited.
In order to avoid size limitation, we apply the second-order perturbation to $\hat{H}_{\mathrm{mat}}$ by assuming $\Delta \gg U,V \gg t_{\mathrm{eh}},t_{\mathrm{e}},t_{\mathrm{h}}$.
We define the unperturbed Hamiltonian $\hat{H}_0$ and the perturbative Hamiltonian $\hat{H}'$ as follows:
\begin{align}
  &\hat{H}_0 = \hat{H}_{\Delta} + \hat{H}_U + \hat{H}_{V} - \sum_{i=\mathrm{h,e}}\frac{W_i}{2}\hat{N}_i,  \\
  &\hat{H}' = \hat{H}_{\mathrm{mat}} - \hat{H}_0.
\end{align}
$\hat{N}_{\mathrm{h}} \coloneq \sum_{{\bf r}_\mathrm{h}} \hat{n}_{{\bf r}_\mathrm{h}}^{\mathrm{h}}$ and $\hat{N}_{\mathrm{e}} \coloneq \sum_{{\bf r}_{\mathrm{e}}} \hat{n}_{{\bf r}_{\mathrm{e}}}$ denote the numbers of holes and electrons, respectively. The corresponding bandwidths are
$W_{\mathrm{h}}=4t_{\mathrm{h}}$ and $W_{\mathrm{e}} = 8t_{\mathrm{e}}$, obtained from the noninteracting Hamiltonian in the limit $t_{\mathrm{eh}}/\Delta \to 0$.  
The matrix element of the second-order perturbed Hamiltonian of the matter part is given by
\begin{align}
  &\braket{i | \hat{H}_{\mathrm{eff}}| j} = \epsilon_i \delta_{ij} + \braket{i | \hat{H}' | j} \nonumber \\
  & + \frac{1}{2}\braket{i| \hat{H}_{t_{\mathrm{eh}}} \left( \frac{1}{\epsilon_i-\hat{H}_0} + \frac{1}{\epsilon_j-\hat{H}_0} \right)  \hat{H}_{t_{\mathrm{eh}}}| j},\label{eq:Heff}
\end{align}
where both $\ket{i}$ and $\ket{j}$ are simultaneous eigenstates of ($\hat{H}_0, \hat{N}_{\mathrm{h}}, \hat{N}_{\mathrm{e}}$), with energy eigenvalues $\epsilon_i$ and $\epsilon_{j}$, respectively.
Because the effective Hamiltonian \eqref{eq:Heff} conserves the number of electrons and holes, the computational cost to calculate the eigenstates is significantly reduced.
For the case of zero total momentum, the dimensions of subspaces with $N_{\mathrm{h}} = N_{\mathrm{e}} =$ 0, 1, and 2 are $1$, $O(N)$, and $O(N^3)$, respectively. Thus, we can investigate the biexciton eigenstates ($N_{\mathrm{h}} = N_{\mathrm{e}} =2$) for $N \sim 10^2$.

By including light-matter interaction terms, the total Hamiltonian in our calculation is given by
\begin{equation}
  \hat{H}(t) = \hat{H}_{\mathrm{eff}} - S\hat{{\bf j}}\cdot {\bf A}(t).
\end{equation}
Our calculations are restricted to $N_{\mathrm{e}} = N_{\mathrm{h}} \leq 2$, which would be justified for weak light.
We use $\Delta = 60t_{\mathrm{e}}$ and $U = V = 10t_{\mathrm{e}}$, which lie in a parameter region where the second-order perturbation is valid. 
We set $t_{\mathrm{h}} = 2t_{\mathrm{e}}$, at which the bandwidths of the electrons and holes are equal ($W_{\mathrm{h}} = W_{\mathrm{e}}$), and $t_{\mathrm{eh}} = t_{\mathrm{e}}$ for simplicity.
The unit of energy (time) is set to $t_{\mathrm{e}}$ ($\hbar/t_{\mathrm{e}}$).

%% file: DMS_SecIV-B.tex
\subsection{\label{SecIVB}Computational details}
As mentioned in Sec.~\ref{SecIVA}, the light pulses are incident normal to the $xy$ plane, and the in-plane component of the wavevectors is fixed at ${\bf k}_{\parallel} = {\bf 0}$.
The vector potentials of the pump pulse and the probe pulse at $z=0$ are set to
\begin{align}
  {\bf A}_{\mathrm{pu}}(t) &=
    A_{\mathrm{pu},0}\exp\left(-\frac{t^2}{2T_{\mathrm{pu}}^2}\right) \cos(\omega_{\mathrm{pu}}t) {\bf e}_x,\\
  {\bf A}_{\mathrm{pr}}(t,\tau) &=
  A_{\mathrm{pr},0}\exp\left(-\frac{(t-\tau)^2}{2T_{\mathrm{pr}}^2}\right)\nonumber \\
  &\hspace{1em}\times 
  \sum_{\mu = x,y}
  e_{\mathrm{pr}}^\mu\cos(\omega_{\mathrm{pr}}(t-\tau)+\theta_{\mathrm{pr}}^\mu){\bf e}_\mu ,
\end{align}
where $(e_{\mathrm{pr}}^x)^2 + (e_{\mathrm{pr}}^y)^2 = 1$.
Here, $\omega_{\mathrm{pu}}$ ($\omega_{\mathrm{pr}}$) is the central frequency of the pump (probe) pulse, and $T_{\mathrm{pu}}$ ($T_{\mathrm{pr}}$) determines the pulse width of the pump (probe) pulse. The parameters $e_{\mathrm{pr}}^{x,y}$ and $\theta_{\mathrm{pr}}^{x,y}$ characterize the polarization of the probe pulse, and their values are listed at the end of this subsection.

After setting the parameters of the light pulses, we calculate the time evolution of the system. The initial state is set to $\ket{\psi(t=-10T_{\mathrm{pu}})} = \ket{0}$, where $\ket{0}$ is the vacuum state of the excitons and also the ground state of $\hat{H}_{\mathrm{eff}}$.
We employ the time-dependent Lanczos method~\cite{Lanczos}. 
The time evolution of the state is computed as
\begin{equation}
  \ket{\psi(t+\delta t/2)} = \sum_{m=1}^M e^{-i\tilde{\epsilon}_m \delta t} \ket{\tilde{\phi}_m(t)} \braket{\tilde{\phi}_m(t) | \psi(t-\delta t/2)},
\end{equation}
where $\ket{\tilde{\phi}_m(t)}$ and $\tilde{\epsilon}_m(t)$ are eigenstates and the corresponding eigenvalues of $\hat{H}(t)$, respectively, calculated using the Lanczos method with the $M$-dimensional Krylov subspace generated from the initial vector $\ket{\psi(t-\delta t/2)}$.
In the time-dependent Lanczos method, we set $\delta t = 0.005\hbar/t_{\mathrm{e}}$ and $M = 10$.
Full reorthogonalization of the Lanczos vectors is adopted during the Lanczos iterations.

For polarization tomography, we use the polarization basis \{$H$, $V$, $D$, $R$\}, where $H$, $V$, $D$, and $R$ denote, respectively, the horizontal polarization, the vertical polarization, the diagonal polarization, and the right-circular polarization.
The diagonal and right-circular polarizations are defined as $\ket{D} = (\ket{H}+\ket{V})/\sqrt{2}$ and $\ket{R} = (\ket{H} -i\ket{V})/\sqrt{2}$.
In our calculation, $H$ $(V)$ is set to $x$ ($y$) polarization, and $\ket{H}$ ($\ket{V}$) is also denoted as $\ket{x}$ ($\ket{y}$).
We have numerically checked that the results of the tomography are unchanged when using other polarization bases.

The detailed procedure of the tomography for the delay time $\tau$ is as follows.
(1) Calculate the time evolution of the state under (i) the pump pulse only and (ii) the probe pulse only, and denote the resulting currents as ${\bf j}_{\mathrm{pu}}(t)$ and ${\bf j}_{\mathrm{pr}}(t,\tau)$, respectively.
The polarization of the probe pulse is denoted by $\alpha$.
Here, the parameters are given by
$(e_{\mathrm{pr}}^x, \theta_{\mathrm{pr}}^x, e_{\mathrm{pr}}^y, \theta_{\mathrm{pr}}^y) = (1,0,0,0)$ for $\alpha = H$, $(0,0,1,0)$ for $\alpha = V$, $(1/\sqrt{2},0,1/\sqrt{2},0)$ for $\alpha = D$, and $(1/\sqrt{2},0,1/\sqrt{2},\pi/2)$ for $\alpha = R$.
(2) Calculate the time evolution of the state when both the pump pulse and the probe pulse are applied and the polarization of the probe pulse is set to $\alpha$ ($\alpha = H, V, D, R$), and compute the expectation value of the current ${\bf j}(t,\tau)$.
(3) Perform the Fourier transform of $\delta {\bf j}(t,\tau) \coloneq {\bf j}(t,\tau) - {\bf j}_{\mathrm{pu}}(t) - {\bf j}_{\mathrm{pr}}(t,\tau)$ with respect to $t$, and obtain the intensity of the $\beta$-polarization ($\beta = H, V, D, R$) component of $\delta {\bf j}(\omega,\tau)$ for $\omega = \omega_{\mathrm{tom}} \sim 2\omega_{\mathrm{pu}} - \omega_{\mathrm{pr}}$ (FWM-related current), $I_{\alpha \beta}(\tau) \coloneq |{\bf e}_{\beta} \cdot\delta{\bf j}(\omega_{\mathrm{tom}},\tau)|^2$.
Here, ${\bf e}_\beta$ is given by ${\bf e}_\beta = {\bf e}_x$ for $\beta = H$, ${\bf e}_y$ for $\beta = V$, $({\bf e}_x + {\bf e}_y)/\sqrt{2}$ for $\beta = D$, and $({\bf e}_x -i {\bf e}_y)/\sqrt{2}$ for $\beta = R$. 
We obtain all $4^2 = 16$ combinations of $I_{\alpha\beta}(\tau)$. 
(4) Perform two-qubit tomography to reconstruct the density matrix $\hat{\rho}_{\mathrm{DMS}}(\tau)$ by replacing $n_{\alpha \beta}$ in Ref.~\cite{Tomography} with $I_{\alpha \beta}$, where $n_{\alpha \beta}$ denotes the number of photon pairs with polarizations $\alpha$ and $\beta$.
The explicit form of $\hat{\rho}_{\mathrm{DMS}}(\tau)$ is given by
\begin{equation}
    \hat{\rho}_{\mathrm{DMS}}(\tau) = \frac{1}{4} \sum_{i_1,i_2=0}^3 \frac{S_{i_1 i_2}(\tau)}{S_{00}(\tau)} \hat{\sigma}_{i_1} \otimes \hat{\sigma}_{i_2} ,
    \label{eq:sec4_tomography}
\end{equation}
where $\hat{\sigma}_0$ is the identity operator and $\hat{\sigma}_{1,2,3}$ are Pauli operators, defined as $\hat{\sigma}_0 = \ket{H}\bra{H} + \ket{V}\bra{V}$, $\hat{\sigma}_1 = \ket{V}\bra{H} + \ket{H}\bra{V}$, $\hat{\sigma}_2 = i\left( \ket{V}\bra{H} - \ket{H}\bra{V} \right)$, and $\hat{\sigma}_3 = \ket{H}\bra{H} - \ket{V}\bra{V}$.
The Stokes parameters $S_{i_1 i_2}(\tau) $ can be obtained from $\{I_{j_1j_2}(\tau) \}$, where $j = 0,1,2,3$ corresponds to $H,V,D,R$, respectively:
\begin{equation}
S_{i_1 i_2}(\tau) = \sum_{j_1,j_2 = 0}^3 (\Upsilon^{-1})_{i_1 j_1} (\Upsilon^{-1})_{i_2 j_2} I_{j_1 j_2}(\tau) .
\end{equation}
The matrix $\Upsilon$ is defined through $\hat{\mu}_i \equiv \ket{i}\bra{i} =  \sum_{j=0}^3 \Upsilon_{ij}\hat{\sigma}_j $.
For example, for $i=3$ (corresponding to $R$), $\ket{R}\bra{R} = \hat{\sigma}_0/2 - \hat{\sigma}_2/2$, i.e., $\Upsilon_{30} = 1/2, \Upsilon_{32} = -1/2, \Upsilon_{31} = \Upsilon_{33} = 0$.
The matrix $\Upsilon$ is given by 
\begin{equation}
\Upsilon = \frac{1}{2} \begin{bmatrix}
1 & 0 & 0 & 1 \\
1 & 0 & 0 & -1 \\
1 & 1 & 0 & 0 \\
1 & 0 & -1 & 0  \end{bmatrix}.
\end{equation}
The normalization by $S_{00}(\tau)$ in Eq.~(\ref{eq:sec4_tomography}) ensures $\mathrm{Tr}\,\hat{\rho}_{\mathrm{DMS}}(\tau) = 1$.

%% file: DMS_SecIV-C.tex
\subsection{\label{SecIVC}Linear and nonlinear optical absorption spectra}
To determine the parameters of the pump and probe pulses, we calculate the one-photon absorption (1PA) and the two-photon absorption (2PA) spectra of $x$-polarized light.
Using eigenstates $\ket{n}$ and eigenenergies $\epsilon_n$ of $\hat{H}_{\mathrm{eff}}$ (i.e., $\hat{H}_{\mathrm{eff}}\ket{n}=\epsilon_n\ket{n}$), the 1PA spectrum is defined as
\begin{equation}
  A_{1\mathrm{PA}}(\omega) \coloneq S\sum_{n}
  \left| \braket{n | \hat{j}^x | 0} \right|^2 \delta(\hbar \omega + \epsilon_0 - \epsilon_n),
\end{equation}
which captures one-photon-allowed exciton states.
Here, $\hat{j}^{\alpha}$ ($\alpha=x,y$) denotes the $\alpha$ component of the current operator $\hat{\bf j}$.
The 2PA spectrum is defined as
\begin{align}
  A_{2\mathrm{PA}}(\omega) \coloneq S^3 \sum_n
  \left|
  \sum_{m} \frac{\braket{n|\hat{j}^x|m} \braket{m|\hat{j}^x|0} }{\hbar \omega + \epsilon_0 - \epsilon_m}
  \right|^2 \delta(2\hbar \omega + \epsilon_0 - \epsilon_n),
\end{align}
which captures two-photon-allowed biexciton eigenstates.
For the 1PA spectrum and for the intermediate states entering the 2PA spectrum, we perform full diagonalization of $\hat{H}_{\mathrm{eff}}$ within the $N_{\mathrm{e}}=N_{\mathrm{h}}=1$ subspace with zero total momentum.
For the final biexciton states in the low-$\omega$ region of the 2PA spectrum, we calculate the low-energy eigenstates of $\hat{H}_{\mathrm{eff}}$ within the subspace $N_{\mathrm{e}} = N_{\mathrm{h}} = 2$ using the implicitly restarted Lanczos method~\cite{resLanczos} implemented in \texttt{scipy.sparse.linalg.eigsh}~\cite{Scipy}.

\begin{figure}[tb]
\includegraphics[width=8cm]{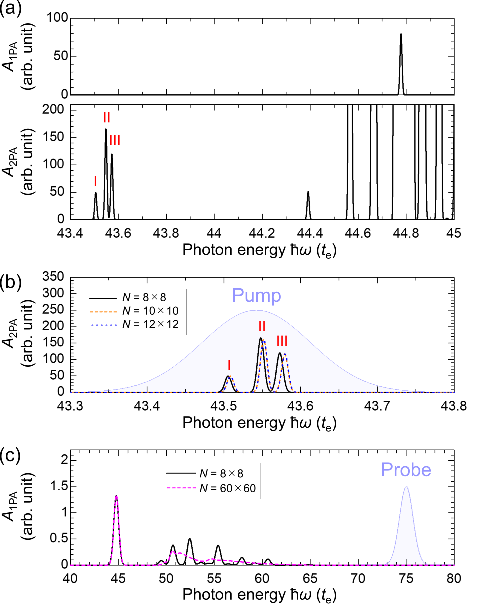}	
\caption{\label{fig:spect} (a) 1PA and 2PA spectra for $N=8\times 8$ sites. Delta functions are convoluted with a Gaussian function with full-width at half-maximum (FWHM) of $0.01t_{\mathrm{e}}$ for visualization. (b) Power spectrum of the pump pulse and 2PA spectra for $N=8\times 8$, $10\times 10$, and $12\times 12$ sites. (c) Power spectrum of the probe pulse and 1PA spectra for $N=8\times 8$ and $60\times 60$.}
\end{figure}

Figure~\ref{fig:spect}(a) shows the 1PA and 2PA spectra for an $8\times 8$-site lattice.
Hereafter, we focus on the lowest energy peak of the 1PA spectrum and the three peaks in the 2PA spectrum located in $43.5t_{\mathrm{e}}<\hbar \omega<43.6t_{\mathrm{e}}$, which are labeled I--III.
The lowest excitonic peak in the 1PA spectrum should be distinguished from the absolute lowest exciton eigenstate.
The latter is a spin-triplet dark exciton, and is therefore absent in the 1PA spectrum.
We have checked that the lowest excitonic peak energy in the 1PA spectrum differs from the lowest exciton energy by only about $10^{-3}t_{\mathrm{e}}$.
Therefore, the 2PA peaks with photon energies lower than that of the lowest 1PA peak correspond to biexciton absorption.
Hereafter, we focus on the biexciton modes I--III.
Figure~\ref{fig:spect}(b) shows the 2PA spectra for $N=8\times 8$, $10\times 10$, and $12 \times 12$.
The 2PA spectra for $N=10\times 10$ and $12 \times 12$ are almost identical.
Compared to these spectra, the spectrum for $N=8\times 8$ is slightly shifted to lower energy.
The amount of the shift is smaller than $0.007t_{\mathrm{e}}$, which is negligible.
Thus, we take $N=8\times 8$ for the simulation of DMS.

Before discussing the main topic, we mention the characteristics of each excited state for the system with $N=8\times 8$ sites.
The lowest excitonic peak in the 1PA spectrum consists of two degenerate exciton eigenstates, distinguished by their
 fourfold rotation eigenvalues $C_4 = \pm i$.
In general, one-photon-allowed excitons have $C_4 = \pm i$.
Similarly, two-photon-allowed biexcitons are characterized by $C_4 = \pm 1$: $C_4$ for biexciton modes I and III is $+1$, and that for mode II is $-1$.
The biexciton modes I--III also have reflection symmetries, $x \to -x$ and $y \to -y$, and the eigenvalue of each mode with respect to both reflection operators is $+1$.
The energies of biexciton modes I, II, and III are $\epsilon_{\mathrm{bi,I}} = 87.01107t_{\mathrm{e}}$, $\epsilon_{\mathrm{bi,II}} = 87.09623t_{\mathrm{e}}$, and $\epsilon_{\mathrm{bi,III}} = 87.14566t_{\mathrm{e}}$, respectively.
The two holes in biexciton modes I--III have antiparallel spins ($h\uparrow$ and $h\downarrow$), and the same is true for the two electrons ($e\uparrow$ and $e\downarrow$).
Strictly speaking, the spin-parallel components ($h\uparrow h\uparrow e\downarrow e\downarrow$ and $h\downarrow h\downarrow e\uparrow e\uparrow$) are nonzero but their weights are less than $10^{-5}$.
Therefore, the four particles constituting the biexciton ($h\uparrow$, $h\downarrow$, $e\uparrow$, $e\downarrow$) are distinguishable, which is an important property for defining the entanglement between two excitons in the biexciton as discussed later.
The average electron-hole distance of the lowest energy exciton is $1.2a$, which reflects that the majority of this state is composed of states in which the electron and hole are next to each other.
The average hole-hole (electron-hole) distances of the biexciton modes I, II, and III are $1.1a$ $(1.2a)$, $1.4a$ $(1.3a)$, and $1.9a$ $(1.3a)$, respectively.
Thus, the electrons and holes in these modes are close to each other.

%% file: DMS_SecIV-D.tex
\subsection{\label{SecIVD}Density matrix of DMS}
Figures~\ref{fig:spect}(b) and \ref{fig:spect}(c) show the power spectra of the pump pulse and the probe pulse, respectively.
The pump pulse is set so as to excite the modes I--III, thereby inducing quantum interference among them.
The probe pulse is off-resonant with the electron-hole system.
The probe pulse remains off-resonant even in a sufficiently large system ($N=60\times 60$), as shown in Fig.~\ref{fig:spect}(c).
The central photon energies of the pump pulse and the probe pulse are set to $\hbar \omega_{\mathrm{pu}} = 43.54t_{\mathrm{e}}$ and $\hbar \omega_{\mathrm{pr}} = 75t_{\mathrm{e}}$, respectively.
The pulse width, defined as the FWHM of the squared envelope of the vector potential, is set to $16.65 \hbar/t_{\mathrm{e}}$ for the pump pulse and $1.665 \hbar/t_{\mathrm{e}}$ for the probe pulse, corresponding to $T_{\mathrm{pu}} = 10\hbar / t_{\mathrm{e}}$ and $T_{\mathrm{pr}} = \hbar / t_{\mathrm{e}}$, respectively.

\begin{figure}[tb]
  \includegraphics[width=8cm]{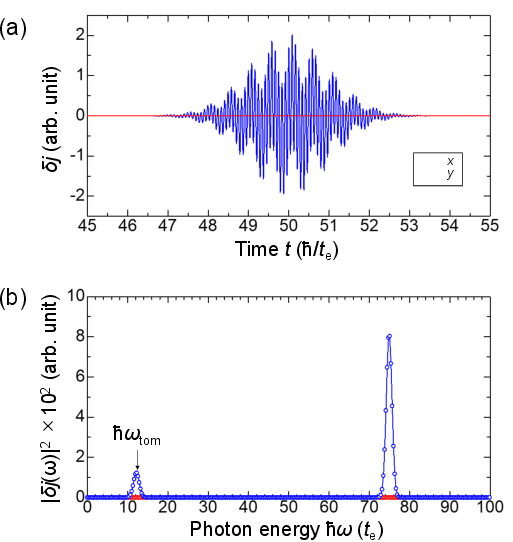}
  \caption{\label{fig:deltaj} (a) Waveforms of $\delta j_x(t,\tau)$ and $\delta j_y(t,\tau)$ for $\tau = 50\hbar/t_{\mathrm{e}}$ for probe polarization $V$. (b) Corresponding power spectra, $|\delta j_x(\omega,\tau)|^2$ and $|\delta j_y(\omega,\tau)|^2$. The photon energy used in the tomography is indicated by a vertical arrow.}
\end{figure}

We focus on $\delta {\bf j}(t,\tau) = \mathbf{j}(t,\tau) - \mathbf{j}_{\mathrm{pu}}(t) - \mathbf{j}_{\mathrm{pr}}(t,\tau)$ where $\mathbf{j}_{\mathrm{pu}}(t)$ ($\mathbf{j}_{\mathrm{pr}}(t,\tau)$) is the current density under the pump (probe) pulse only.
Figure~\ref{fig:deltaj}(a) shows the waveforms of
$\delta j_x(t,\tau)$ and $\delta j_y(t,\tau)$
for $\tau = 50\hbar/t_{\mathrm{e}}$ with the probe polarization set to $V$ (${\bf A}_{\mathrm{pr}} \parallel {\bf e}_y$)
, where $\delta j_x$ vanishes within numerical accuracy.
$\delta {\bf j}(t,\tau)$ is nonzero only while the probe pulse is applied.
This property arises because the probe pulse is off-resonant.
Figure~\ref{fig:deltaj}(b) shows the corresponding power spectra, $|\delta j_x(\omega,\tau)|^2$ and $|\delta j_y(\omega,\tau)|^2$. The latter spectrum has two peaks: the lower and upper peaks located at $\omega = 2\omega_{\mathrm{pu}} - \omega_{\mathrm{pr}}(=12.08 t_{\mathrm{e}}/\hbar)$ and $\omega = \omega_{\mathrm{pr}}$, respectively.
Thus, the oscillatory component of the lower peak corresponds to the FWM signal, and we focus on the polarized intensity of $\delta {\bf j}(\omega_{\mathrm{tom}},\tau)$ at $\hbar \omega_{\mathrm{tom}} = 12.249t_{\mathrm{e}}$ for the tomography, where $\omega_{\mathrm{tom}}$ is chosen near the lower peak.
We conduct the tomography described in Sec.~\ref{SecIVB} for various delay times $\tau$ to obtain the density matrix $\hat{\rho}_{\mathrm{DMS}}(\tau)$.
The purity of the numerically obtained $\hat{\rho}_{\mathrm{DMS}}(\tau)$ is very close to 1: $0.999995 < \mathrm{Tr} \left[\hat{\rho}_{\mathrm{DMS}}(\tau)\right]^2 < 1$, indicating that the deviation of the purity from 1 is due to numerical errors and that the obtained state can be regarded as pure.

\begin{figure}[tb]
  \includegraphics[width=8cm]{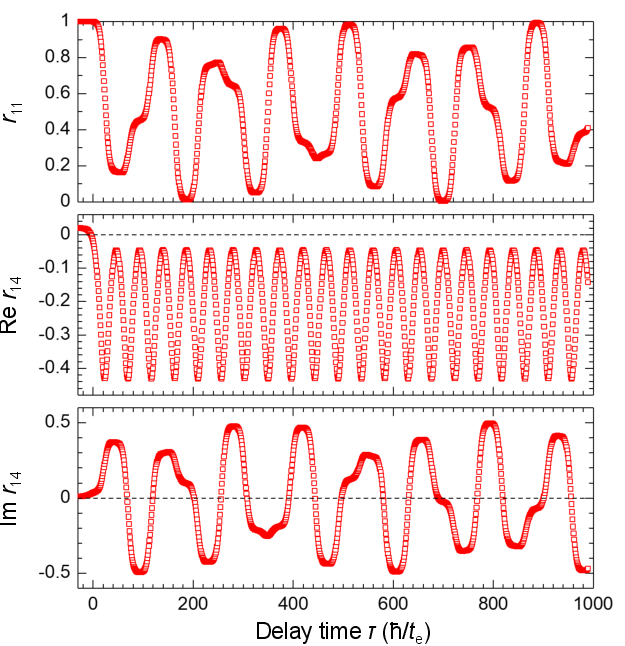}
  \caption{\label{fig:rho_DMS} Time evolution of the matrix elements of $\hat{\rho}_{\mathrm{DMS}}(\tau)$. }
\end{figure}

The obtained $\hat{\rho}_{\mathrm{DMS}}(\tau)$ can be expressed as
\begin{align}
\hat{\rho}_{\mathrm{DMS}}(\tau) 
&=r_{11}(\tau)\ket{HH}\bra{HH} + r_{14}(\tau)\ket{HH}\bra{VV}  \nonumber \\
& \hspace{1em} + r_{41}(\tau)\ket{VV}\bra{HH} + r_{44}(\tau)\ket{VV}\bra{VV}  \nonumber \\
&=
  \begin{bmatrix}
    r_{11}(\tau) & 0 & 0 & r_{14}(\tau) \\
    0 & 0 & 0 & 0 \\
    0 & 0 & 0 & 0 \\
    r_{41}(\tau) & 0 & 0 & r_{44}(\tau)
  \end{bmatrix},
  \label{eq:rho_DMS}
\end{align}
where $r_{44}(\tau) = 1-r_{11}(\tau)$ and $r_{41}(\tau) = r_{14}^*(\tau)$.
Here, the matrix elements $\braket{VH|\hat{\rho}_{\mathrm{DMS}}|\alpha \beta}$, $\braket{HV|\hat{\rho}_{\mathrm{DMS}}|\alpha \beta}$, $\braket{\alpha \beta|\hat{\rho}_{\mathrm{DMS}}|VH}$, and $\braket{\alpha \beta|\hat{\rho}_{\mathrm{DMS}}|HV}$ are zero due to the mirror symmetries of the biexciton modes I--III.
Therefore, the independent matrix elements are $r_{11} \in \mathbb{R}$ and $r_{14} \in \mathbb{C}$, whose time evolutions are shown in Fig.~\ref{fig:rho_DMS}.
The matrix elements oscillate with respect to the delay time $\tau$, whose frequency is related to the difference in energy of the excited biexciton modes.
The oscillation frequencies $\hbar \Omega$ are given by $\hbar \Omega = \epsilon_{\mathrm{bi,III}} -  \epsilon_{\mathrm{bi,I}} = 0.13459t_{\mathrm{e}}$ for $\Re r_{14}$ and $\hbar \Omega = \epsilon_{\mathrm{bi,II}} -  \epsilon_{\mathrm{bi,I}} = 0.08516t_{\mathrm{e}}$, $\epsilon_{\mathrm{bi,III}} -  \epsilon_{\mathrm{bi,II}} = 0.04943t_{\mathrm{e}}$ for $r_{11}$ and $\Im r_{14}$.
Since the reconstructed density matrix $\hat{\rho}_{\mathrm{DMS}}(\tau)$ can be regarded as pure, we define the corresponding state vector $\ket{\psi_{\mathrm{DMS}}(\tau)}$ by
$\hat{\rho}_{\mathrm{DMS}}(\tau)\equiv \ket{\psi_{\mathrm{DMS}}(\tau)}\bra{\psi_{\mathrm{DMS}}(\tau)}$.
In the following, we discuss the entanglement dynamics of $\ket{\psi_{\mathrm{DMS}}(\tau)}$.

%% file: DMS_SecIV-E-1.tex
\subsection{\label{SecIVE}Entanglement dynamics}
\subsubsection{DMS-entanglement and the entanglement in the biexciton}
\begin{figure}[tb]
  \includegraphics[width=8cm]{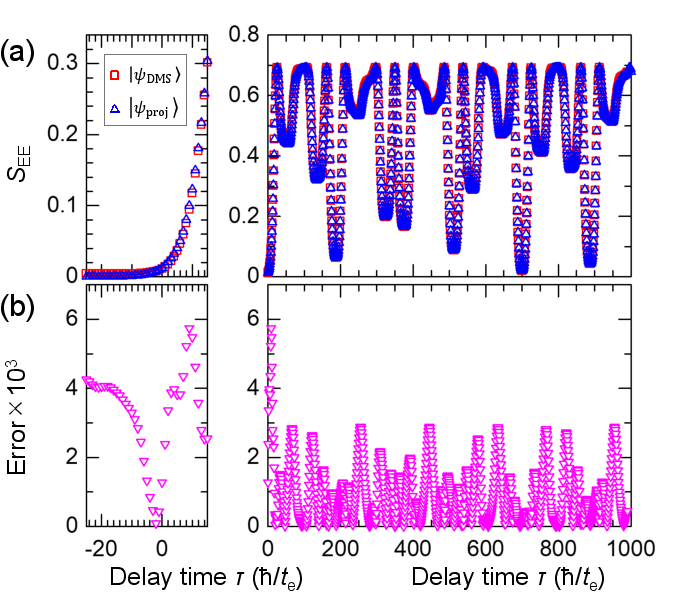}
  \caption{\label{fig:SEE_compare} (a) $\tau$-dependence of $S_{\mathrm{EE}}(\ket{\psi_{\mathrm{DMS}}(\tau)})$ and $S_{\mathrm{EE}}(\ket{\psi_{\mathrm{proj}}(\tau)})$. The definition of $\ket{\psi_{\mathrm{proj}}(\tau)}$ is given by Eq.~(\ref{eq:def_psi_proj}). The range of $\tau$ in the left (right) panel is $-25\hbar/t_{\mathrm{e}} \leq \tau \leq 15\hbar/t_{\mathrm{e}}$ ($0 \leq \tau \leq 1000\hbar/t_{\mathrm{e}}$).
    (b) The difference in EE between $\ket{\psi_{\mathrm{DMS}}(\tau)}$ and $\ket{\psi_{\mathrm{proj}}(\tau)}$.}
\end{figure}

Next, we focus on the dynamics of the entanglement.
We use entanglement entropy (EE) as the measure of the entanglement.
The EE of the state $\ket{\psi}$ is denoted by $S_{\mathrm{EE}}(\ket{\psi})$, where $\ket{\psi}$ is either $\ket{\psi_{\mathrm{DMS}}(\tau)}$ or a state of a biexciton composed of $h\uparrow$, $h\downarrow$, $e\uparrow$, and $e\downarrow$.
In the latter case, we consider the EE between exciton A ($h\uparrow$, $e\downarrow$) and exciton B ($h\downarrow$, $e\uparrow$) as a measure of the biexciton entanglement, which is defined as
\begin{align}
  S_{\mathrm{EE}}(\ket{\psi}) &= -\Tr( \hat{\rho}_{\mathrm{A}} \ln \hat{\rho}_{\mathrm{A}}), \\
   \hat{\rho}_{\mathrm{A}} &\coloneq \Tr_{\mathrm{B}} ( \ket{\psi} \bra{\psi} ).
\end{align}
We define $S_{\mathrm{DMS}}(\tau) \coloneq S_{\mathrm{EE}}(\ket{\psi_{\mathrm{DMS}}}) = -\sum_{i=1,4} r_{ii}\ln r_{ii}$.
Figure~\ref{fig:SEE_compare}~(a) shows $S_{\mathrm{DMS}}(\tau)$, which grows and oscillates with a maximum value of $\ln 2$ after the pump pulse is applied.

$S_{\mathrm{DMS}}$ is compared with the EE of the following two-qubit state,
\begin{align}
  \ket{\psi_{\mathrm{proj}}(\tau)} &\coloneq \sum_{\alpha,\beta = x,y}
  \mathcal{N} \ket{\Phi^{\alpha \beta}} \braket{\Phi^{\alpha \beta} | \psi_{\mathrm{bi}}(\tau)} \nonumber \\
  & \equiv \frac{\hat{P}_{\Phi}\ket{\psi_{\mathrm{bi}}(\tau)}}{\| \hat{P}_{\Phi}\ket{\psi_{\mathrm{bi}}(\tau)} \|},  \\
  \ket{\Phi^{\alpha \beta}} & \coloneq \frac{\hat{j}_{+,\uparrow}^\alpha \hat{j}_{+,\downarrow}^\beta \ket{0}}{ \| \hat{j}_{+,\uparrow}^\alpha \hat{j}_{+,\downarrow}^\beta \ket{0} \| },
  \label{eq:def_psi_proj}
\end{align}
where $\mathcal{N}$ ($>0$) is a normalization factor, 
and $\hat{P}_{\Phi}$ is a projection operator defined as $\hat{P}_{\Phi} \coloneq \sum_{\alpha,\beta=x,y}\ket{\Phi^{\alpha \beta}}\bra{\Phi^{\alpha \beta}}$.
Note that $\{\ket{\Phi^{\alpha \beta}} \}_{\alpha,\beta=x,y}$ is orthonormal: $\braket{\Phi^{\alpha \beta}|\Phi^{\alpha' \beta'}} = \delta_{\alpha \alpha'} \delta_{\beta \beta'}$.
Figure~\ref{fig:SEE_compare}(a) shows the comparison between $S_{\mathrm{DMS}}(\tau)$ and $S_{\mathrm{proj}}(\tau) \coloneq S_{\mathrm{EE}}(\ket{\psi_{\mathrm{proj}}(\tau)})$.
These two EEs differ only slightly.
Figure~\ref{fig:SEE_compare}(b) shows the difference, $|S_{\mathrm{DMS}}(\tau) - S_{\mathrm{proj}}(\tau)|$.
The maximum value of the difference is about $6\times 10^{-3}$, which is about two orders of magnitude smaller than their respective EE.
The reason for this near equality is as follows.
Considering $\psi^{\alpha \beta}_{{\bf k}_{\parallel}={\bf 0}}(\omega=\omega_{\mathrm{tom}},\tau)$ in Sec.~\ref{SecIII}, which is an analytical representation of the $(\alpha,\beta)$-component of $\ket{\psi_{\mathrm{DMS}}(\tau)}$ for $\tau \gg T_{\mathrm{pu}}$, the following approximation can be made:
\begin{align}
  &\psi^{\alpha \beta}_{{\bf k}_{\parallel}={\bf 0}}(\omega_{\mathrm{tom}},\tau) \propto \braket{\phi^{\alpha \beta}_{{\bf k}_{\parallel}={\bf 0}}(\omega_{\mathrm{tom}}) | \psi_{\mathrm{bi}}(\tau)}  \nonumber \\
  &= \braket{0| \hat{j}_-^\beta  \frac{1}{\hbar \omega_{\mathrm{tom}} - \epsilon_{\mathrm{bi}}+(\hat{H}_{\mathrm{eff}}-\epsilon_0 ) +i\eta}\hat{j}_-^\alpha |\psi_{\mathrm{bi}}(\tau)} \nonumber \\
  &\hspace{1em}-\braket{0|  \hat{j}_-^\alpha  \frac{1}{\hbar \omega_{\mathrm{tom}} - (\hat{H}_{\mathrm{eff}}-\epsilon_0 ) +i\eta} \hat{j}_-^\beta | \psi_{\mathrm{bi}}(\tau)} \nonumber \\
    &\simeq \braket{0| \hat{j}_-^\beta  \frac{1}{\hbar \omega_{\mathrm{tom}} - \epsilon_{\mathrm{bi}}+\epsilon_{\mathrm{ex}}}\hat{j}_-^\alpha |\psi_{\mathrm{bi}}(\tau)} \nonumber \\
  &\hspace{1em}-\braket{0|  \hat{j}_-^\alpha  \frac{1}{\hbar \omega_{\mathrm{tom}} - \epsilon_{\mathrm{ex}}} \hat{j}_-^\beta | \psi_{\mathrm{bi}}(\tau)}.
\end{align}
Here, we apply an approximation that replaces $\hat{H}_{\mathrm{eff}}-\epsilon_0 \pm i\eta$ in the denominators with $\epsilon_{\mathrm{ex}}$, which is the lowest energy of the one-photon-allowed exciton.
This approximation may be justified when both $\ket{\psi_{\mathrm{bi}}}$ and the lowest-energy one-photon-allowed exciton show a localized nature and $\omega_{\mathrm{tom}}$ is away from the pole of the fraction.
Because $\hat{j}_-^\alpha$ and $\hat{j}_-^\beta$ commute, $\braket{ \phi^{\alpha \beta}_{{\bf k}_{\parallel}={\bf 0}}(\omega_{\mathrm{tom}},\tau)  | \psi_{\mathrm{bi}}(\tau)}  \propto \braket{0|\hat{j}_{-}^\beta \hat{j}_{-}^\alpha |\psi_{\mathrm{bi}}(\tau)}$ can be obtained.
Recalling that $\ket{\psi_{\mathrm{bi}}(\tau)}$ consists of two holes ($h\uparrow, h\downarrow$) and two electrons ($e\uparrow, e\downarrow$), one finds that 
\begin{align}
  \braket{0|\hat{j}_{-}^\beta \hat{j}_{-}^\alpha |\psi_{\mathrm{bi}}(\tau)}
  &= \braket{0|\hat{j}^\beta_{-,\downarrow}  \hat{j}^\alpha_{-,\uparrow}|\psi_{\mathrm{bi}}(\tau)} \nonumber \\
  &\hspace{1em}+ \braket{0|\hat{j}^\beta_{-,\uparrow}  \hat{j}^{\alpha}_{-,\downarrow}|\psi_{\mathrm{bi}}(\tau)}.
\end{align}
Here, we consider the symmetric transformation of reversing all spins of electrons and holes.
The eigenvalue of this symmetric transformation is $+1$ for two-photon-allowed eigenstates of $\hat{H}_{\mathrm{eff}}$.
Thus, it follows approximately that
\begin{equation}
  \psi^{\alpha \beta}_{{\bf k}_{\parallel}={\bf 0}}(\omega_{\mathrm{tom}},\tau) \propto \braket{0|\hat{j}^\beta_{-,\downarrow}  \hat{j}^\alpha_{-,\uparrow}|\psi_{\mathrm{bi}}(\tau)} = \braket{\Phi^{\alpha \beta}|\psi_{\mathrm{bi}}(\tau)}.
\end{equation}
Therefore, $\braket{\alpha \beta |  \psi_{\mathrm{DMS}}(\tau)}$ is almost equal to $\braket{\alpha \beta| \psi_{\mathrm{proj}}(\tau)}$, and hence $S_{\mathrm{DMS}}(\tau)$ is almost equal to $S_{\mathrm{proj}}(\tau)$.
Hereafter, we use $S_{\mathrm{proj}}(\tau)$ as the EE that is accessible via DMS and compare it with the EE of the biexciton.

%% file: DMS_SecIV-E-2.tex
\begin{figure}[tb]
  \includegraphics[width=8cm]{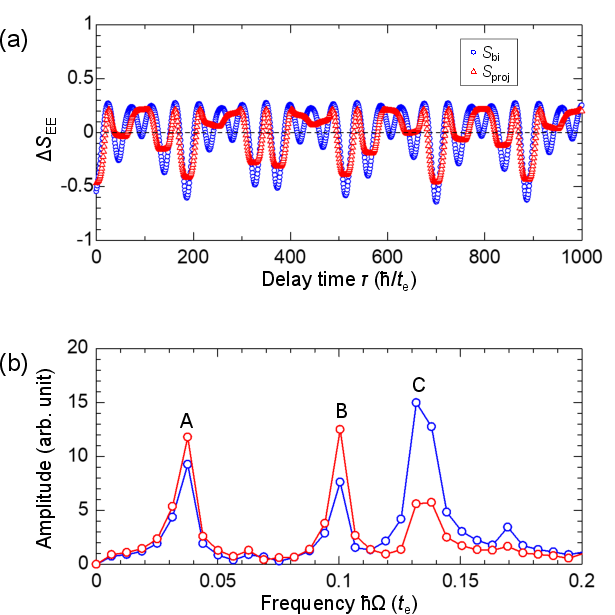}
  \caption{\label{fig:Sbi_Sproj} (a) Entanglement dynamics of $\ket{\psi_{\mathrm{bi}}(\tau)}$ and $\ket{\psi_{\mathrm{proj}}(\tau)}$. (b) Corresponding amplitude spectra of EEs shown in (a). $S_{\mathrm{bi}}$ ($S_{\mathrm{proj}}$) is shown in blue (red).}
\end{figure}

Next, we consider the EE of the biexciton, $S_{\mathrm{bi}}(\tau) \coloneq S_{\mathrm{EE}}(\ket{\psi_{\mathrm{bi}}(\tau)})$, and compare it with $S_{\mathrm{proj}}(\tau)$.
We decompose the EE into a time-averaged component and an oscillatory component: $  S_{\mathrm{EE}}(\tau)  = \bar{S}_{\mathrm{EE}} + \Delta S_{\mathrm{EE}}(\tau)$, where the subscript ``EE'' denotes, for example, ``bi'' or ``proj''. The time-averaged component, $\bar{S}_{\mathrm{EE}}$, is defined as
\begin{equation}
  \bar{S}_{\mathrm{EE}} \coloneq \frac{1}{  \tau_{\mathrm{max}}  }\int^{\tau_{\mathrm{max}} }_{0} \mathrm{d}\tau \  S_{\mathrm{EE}}(\tau).
\end{equation}
We set $\tau_{\mathrm{max}} = 1000\hbar/t_{\mathrm{e}}$ unless otherwise noted. 
Although the average values of the EE differ significantly ($\bar{S}_{\mathrm{bi}} = 3.717$ and $\bar{S}_{\mathrm{proj}} = 0.480$), the oscillatory components of the EE, $\Delta S_{\mathrm{bi}}$ and $\Delta S_{\mathrm{proj}}$, are rather similar, as shown in Fig.~\ref{fig:Sbi_Sproj}(a).

In order to evaluate the similarity of these dynamics, we introduce the cosine similarity between $\Delta f_1(\tau)$ and $\Delta f_2(\tau)$, defined as
\begin{equation}
  \cos \theta \coloneq \frac{\int^{\tau_{\mathrm{max}}}_0 \mathrm{d}\tau\, \Delta f_1(\tau) \Delta f_2(\tau)}{\sqrt{\int^{\tau_{\mathrm{max}}}_0 \mathrm{d}\tau \ [\Delta f_1(\tau)]^2} \sqrt{\int^{\tau_{\mathrm{max}}}_0 \mathrm{d}\tau \ [\Delta f_2(\tau)]^2}}
\end{equation}
The cosine similarity
takes its maximum ($+1$) or minimum ($-1$) value 
when the two oscillatory waveforms are proportional.
Although the cosine similarity depends on $\tau_{\mathrm{max}}$, the $\tau_{\mathrm{max}}$-dependence becomes weaker as $\tau_{\mathrm{max}}$ increases.
The calculated cosine similarity between $\Delta S_{\mathrm{bi}}$ and $\Delta S_{\mathrm{proj}}$ is $0.80 < \cos \theta < 0.83$ for $700\hbar/t_{\mathrm{e}} < \tau < 1000\hbar/t_{\mathrm{e}}$, which indicates that the two oscillatory waveforms have a correlation.
Hereafter, we focus on the dynamics of $\Delta S_{\mathrm{EE}}(\tau)$.

Figure~\ref{fig:Sbi_Sproj}~(b) shows the amplitude spectra of $\Delta S_{\mathrm{bi}}$ and $\Delta S_{\mathrm{proj}}$.
The frequency positions with large amplitudes are common in these spectra, labeled A, B, and C, and their angular frequencies multiplied by $\hbar$ are $\hbar \Omega_{\mathrm{A}} = -\epsilon_{\mathrm{bi,I}} + 2\epsilon_{\mathrm{bi,II}} - \epsilon_{\mathrm{bi,III}} = 0.03574t_{\mathrm{e}}$, $\hbar \Omega_{\mathrm{B}} = -2\epsilon_{\mathrm{bi,II}} + 2\epsilon_{\mathrm{bi,III}} = 0.09885t_{\mathrm{e}}$, $\hbar \Omega_{\mathrm{C}} = -\epsilon_{\mathrm{bi,I}} + \epsilon_{\mathrm{bi,III}} = 0.13459t_{\mathrm{e}}$, respectively.
There is a large difference in the amplitudes of $\Delta S_{\mathrm{bi}}$ and $\Delta S_{\mathrm{proj}}$ at $\Omega = \Omega_{\mathrm{C}}$, which is related to entanglement with respect to the exciton wavevector, as discussed later.

%% file: DMS_SecIV-E-3.tex
\subsubsection{Analysis of the entanglement focusing on the wavevectors of the excitons}
We focus on the wavevectors of the biexciton and its two constituent excitons.
Because the wavevectors of the pump pulse projected onto the $xy$-plane are ${\bf k}_{\mathrm{pu},\parallel} = \mathbf{0}$, the two-dimensional wavevector of the biexciton is $2{\bf k}_{\mathrm{pu},\parallel} = {\bf 0}$.
Consequently, the two-dimensional wavevectors of exciton A ($h\uparrow$, $e\downarrow$) and exciton B ($h\downarrow$, $e\uparrow$) are denoted by ${\bf k}_{\mathrm{A}}$ and ${\bf k}_{\mathrm{B}}$, respectively, and ${\bf k}_{\mathrm{A}} + {\bf k}_{\mathrm{B}} = 2{\bf k}_{\mathrm{pu},\parallel} = {\bf 0}$ holds.
Although there are many combinations of ${\bf k}_{\mathrm{A}}$ and ${\bf k}_{\mathrm{B}}$ in the biexciton state, only the component with ${\bf k}_{\mathrm{A}} = {\bf k}_{\mathrm{B}} = {\bf 0}$ is related to DMS in this simulation because the in-plane wavevectors of both the probe pulse and the generated FWM light are ${\bf 0}$.

\begin{figure}[tb]
  \includegraphics[width=8cm]{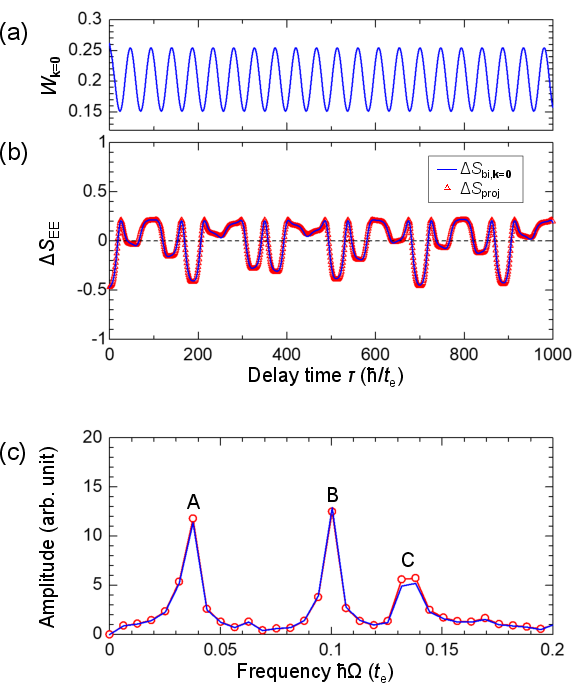}
  \caption{\label{fig:Sbik0_Sproj} (a) $\tau$-dependence of $W_{{\bf k}={\bf 0}}(\tau)$. (b) Entanglement dynamics of $\ket{\psi_{\mathrm{bi},{\bf k}={\bf 0}}(\tau)}$ and $\ket{\psi_{\mathrm{proj}}(\tau)}$. (c) Amplitude spectra corresponding to the dynamics of EEs shown in (b). In (b) and (c), the data for $\Delta S_{\mathrm{bi},{\bf k}={\bf 0}}$ ($\Delta S_{\mathrm{proj}}$) are shown in blue (red).}
\end{figure}

Here, we extract the component with ${\bf k}_{\mathrm{A}} = {\bf k}_{\mathrm{B}} = {\bf 0}$ from the biexciton state $\ket{\psi_{\mathrm{bi}}(\tau)}$.
The basis of exciton A (exciton B) can be expressed as $\ket{{\bf k}_{\mathrm{A}},j_{\mathrm{A}}}$ ($\ket{{\bf k}_{\mathrm{B}},j_{\mathrm{B}}}$), which is an eigenstate of the translational operator $\hat{T}_{\bf R}$ with the corresponding eigenvalue $e^{-i {\bf k}_{\mathrm{A}} \cdot {\bf R} }$ ($e^{-i {\bf k}_{\mathrm{B}} \cdot {\bf R} }$). The indices $j_{\mathrm{A}}$ and $j_{\mathrm{B}}$ represent the degrees of freedom other than the wavevector.
The basis of the biexciton state can be expressed as $\ket{{\bf k}_{\mathrm{A}},j_{\mathrm{A}}} \otimes \ket{{\bf k}_{\mathrm{B}},j_{\mathrm{B}}}$.
Because ${\bf k}_{\mathrm{A}} + {\bf k}_{\mathrm{B}} = {\bf 0}$, the biexciton state $\ket{\psi_{\mathrm{bi}}(\tau)}$ takes the form
\begin{align}
  \ket{\psi_{\mathrm{bi}}(\tau)} &= \sum_{{\bf k}j_{\mathrm{A}} j_{\mathrm{B}}} \psi_{{\bf k}j_{\mathrm{A}} j_{\mathrm{B}}}(\tau) \ket{{\bf k},j_{\mathrm{A}}} \otimes \ket{-{\bf k},j_{\mathrm{B}}} \nonumber \\
  &\equiv \sum_{{\bf k}}\sqrt{W_{{\bf k}}(\tau)}\ket{\psi_{\mathrm{bi},{\bf k}}(\tau)}.
  \label{eq:psi_bi_k}
\end{align}
Here, $\ket{\psi_{\mathrm{bi},{\bf k}}(\tau)}$ and $W_{{\bf k}}(\tau)$ are the component of ${\bf k}_{\mathrm{A}} = -{\bf k}_{\mathrm{B}} \equiv {\bf k}$ and the corresponding weight, respectively, where $\ket{\psi_{\mathrm{bi},{\bf k}}(\tau)}$ is normalized ($\braket{\psi_{\mathrm{bi},{\bf k}}(\tau) | \psi_{\mathrm{bi},{\bf k}}(\tau)} = 1$).
We can extract the component of ${\bf k}_{\mathrm{A}} = -{\bf k}_{\mathrm{B}} = {\bf k}$ using the formula:
\begin{equation}
   \sqrt{W_{{\bf k}}(\tau)}\ket{\psi_{\mathrm{bi},{\bf k}}(\tau)} = \frac{1}{N/2}\sum_{{\bf R}}\hat{T}_{{\bf R},\mathrm{A}}e^{i{\bf k}\cdot {\bf R}}\ket{\psi_{\mathrm{bi}}(\tau)},
  \label{eq:calc_psi_bi_k}
\end{equation}
where $\hat{T}_{{\bf R},\mathrm{A}}$ is the translational operator which acts on the basis of exciton A but does not act on that of exciton B.
The sum over $\mathbf{R}$ in the above formula runs over non-equivalent lattice vectors, the number of which is $N/2$.
$W_{{\bf k}}$ is maximized at ${\bf k} = {\bf 0}$ in this calculation.
Figure~\ref{fig:Sbik0_Sproj}(a) shows the oscillatory dynamics of $W_{{\bf k}={\bf 0}}(\tau)$ with the angular frequency  $\Omega_{\mathrm{C}}$.
The time average of $W_{{\bf k}={\bf 0}}(\tau)$ is 0.203, while the maximum time-averaged value of $W_{{\bf k}\neq{\bf 0}}(\tau)$ is found to be 0.083.
We denote $S_{\mathrm{EE}}(\ket{\psi_{\mathrm{bi},{\bf k}}(\tau)})$ as $S_{\mathrm{bi},{\bf k}}(\tau)$ and compare $S_{\mathrm{bi},{\bf k}={\bf 0}}(\tau)$ with $S_{\mathrm{proj}}(\tau)$.
Their time averages are $\bar{S}_{\mathrm{bi},{\bf k}={\bf 0}} = 0.750$ and $\bar{S}_{\mathrm{proj}} = 0.480$.
Figure~\ref{fig:Sbik0_Sproj}(b) shows the dynamics of the EEs, $\Delta S_{\mathrm{bi},{\bf k}={\bf 0}}(\tau)$ and $\Delta S_{\mathrm{proj}}(\tau)$, where $\Delta S_{\mathrm{bi},{\bf k}={\bf 0}}(\tau) = S_{\mathrm{bi},\mathbf{k} ={\bf 0}}(\tau) - \bar{S}_{\mathrm{bi},{\bf k}={\bf 0}}$.
These entanglement dynamics are almost identical.
The cosine similarity between $\Delta S_{\mathrm{bi},{\bf k}={\bf 0}}(\tau)$ and $\Delta S_{\mathrm{proj}}(\tau)$ is $0.99858 < \cos \theta < 0.99878$ for $700\hbar /t_{\mathrm{e}} \leq \tau_{\mathrm{max}} \leq 1000\hbar /t_{\mathrm{e}}$.
The amplitude spectra of these EEs are also very similar in all frequency regions, as shown in Fig.~\ref{fig:Sbik0_Sproj}(c).
This result demonstrates that DMS enables us to investigate the entanglement dynamics of
$\ket{\psi_{\mathrm{bi},{\bf k}={\bf 0}}(\tau)}$.

\begin{figure}[tb]
  \includegraphics[width=8cm]{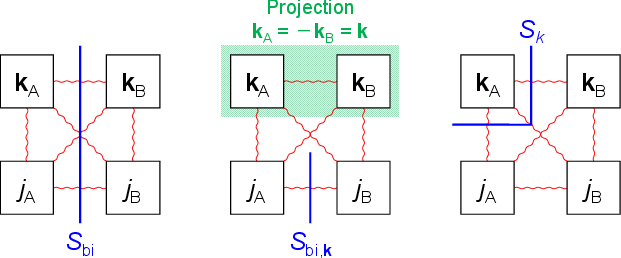}
  \caption{\label{fig:partition} Schematic of the partitions of the system considered when calculating EEs. Degrees of freedom of the biexciton can be divided into ${\bf k}_{\mathrm{A}}$, $j_{\mathrm{A}}$, ${\bf k}_{\mathrm{B}}$, and $j_{\mathrm{B}}$ as mentioned in the main text. Red wavy lines indicate entanglement between each degree of freedom, and blue lines indicate the partition. }
\end{figure}

\begin{figure}[tb]
  \includegraphics[width=8cm]{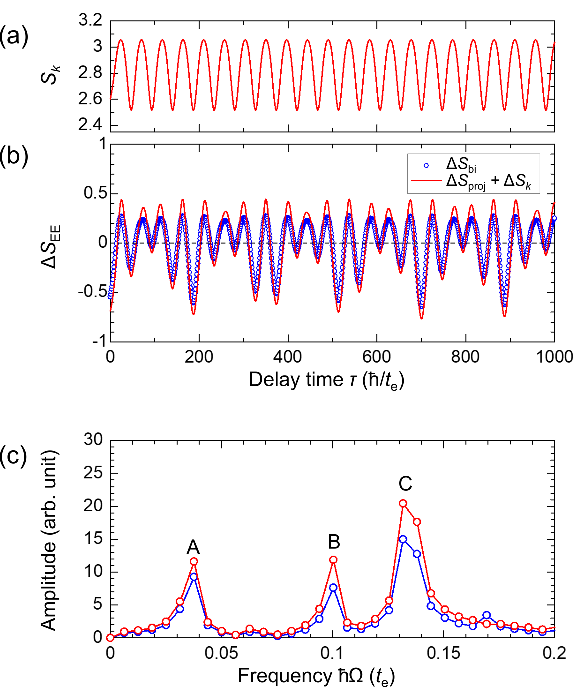}
  \caption{\label{fig:Sbi_SprojplusSk} (a) $\tau$-dependence of the entanglement in terms of the wavevector, $S_k(\tau)$, defined in the main text. (b) Entanglement dynamics of $\Delta S_{\mathrm{bi}}(\tau)$ and $\Delta S_{\mathrm{proj}} + \Delta S_k$. (c) Amplitude spectra corresponding to the entanglement dynamics shown in (b). In (b) and (c), the data for $\Delta S_{\mathrm{bi}}$ ($\Delta S_{\mathrm{proj}} + \Delta S_k$) are shown in blue (red).}
\end{figure}

Next, we consider the relation between $S_{\mathrm{bi},{\bf k}}$ and $S_{\mathrm{bi}}$.
Using the constraint ${\bf k}_{\mathrm{A}} + {\bf k}_{\mathrm{B}} = {\bf 0}$, $S_{\mathrm{bi}}$ can be decomposed as follows:
\begin{equation}
  S_{\mathrm{bi}}(\tau) = \sum_{{\bf k}} W_{{\bf k}}(\tau)S_{\mathrm{bi},{\bf k}}(\tau) + S_k(\tau).
  \label{eq:Sbi_decomposition}
\end{equation}
The derivation of this formula is found in Appendix~\ref{AppendixD}.
Here, $S_k(\tau)$ is the EE between ${\bf k}_{\mathrm{A}}$ and the remaining degrees of freedom ($j_{\mathrm{A}}$, ${\bf k}_{\mathrm{B}}$, and $j_{\mathrm{B}}$). The corresponding density matrix is defined as
\begin{align}
      & \hat{\rho}_{{\bf k}_{\mathrm{A}}}(\tau) \coloneq \Tr_{{\bf k}_{\mathrm{B}} j_{\mathrm{A}} j_{\mathrm{B}} } \left[\ket{\psi_{\mathrm{bi}}(\tau)} \bra{\psi_{\mathrm{bi}}(\tau)}  \right],
\end{align}
and $S_{k}(\tau)$ is given by
\begin{align}
  &S_k(\tau) = -\Tr \left[ \hat{\rho}_{{\bf k}_{\mathrm{A}}}(\tau) \ln \hat{\rho}_{{\bf k}_{\mathrm{A}}}(\tau) \right] = -\sum_{{\bf k}}W_{\bf k}(\tau) \ln W_{\bf k}(\tau).
\end{align}
The second expression corresponds to the Shannon entropy of $\{ W_{{\bf k}}(\tau) \}$, characterizing the entanglement of the wave vector.
Figure~\ref{fig:partition} shows a schematic of the partitions of the system considered when calculating the EEs ($S_{\mathrm{bi}}$, $S_{\mathrm{bi},{\bf k}}$, and $S_k$).
Figure~\ref{fig:Sbi_SprojplusSk}(a) shows the $\tau$ dependence of $S_k(\tau)$.
The time average of $S_k(\tau)$, denoted by $\bar{S}_k$, is $2.824$.
$S_k(\tau)$ oscillates at angular frequencies $n\Omega_{\mathrm{C}}$ with $n\in\mathbb{N}$, and the angular frequency of the main oscillatory component is $\Omega_{\mathrm{C}}$ ($n=1$).

If there exist $\psi_{{\bf k}}$ and $\psi_{ j_{\mathrm{A}} j_{\mathrm{B}} } $ such that $\psi_{{\bf k} j_{\mathrm{A}} j_{\mathrm{B}}} = \psi_{{\bf k}} \psi_{ j_{\mathrm{A}} j_{\mathrm{B}}}$ (separable), $S_{\mathrm{bi},{\bf k}}$ does not depend on ${\bf k}$ and $S_{\mathrm{bi}} = S_{\mathrm{bi},{\bf k}={\bf 0}} + S_k$ holds.
Even when $\psi_{{\bf k} j_{\mathrm{A}} j_{\mathrm{B}}}$ cannot be expressed as $\psi_{{\bf k}} \psi_{ j_{\mathrm{A}} j_{\mathrm{B}}}$, $S_{\mathrm{bi}}$ can be approximated by $S_{\mathrm{bi},{\bf k}={\bf 0}} + S_k$ if $W_{{\bf k}}(S_{\mathrm{bi},{\bf k}} - S_{\mathrm{bi},{\bf k}={\bf 0}})$ is small.
Here, we compare $S_{\mathrm{bi}}(\tau)$ and $S_{\mathrm{bi},{\bf k}={\bf 0}}(\tau) + S_k(\tau)$.
The time averages of $S_{\mathrm{bi}}(\tau)$ and $S_{\mathrm{bi},{\bf k}={\bf 0}}(\tau) + S_k(\tau)$ are 3.717 and 3.574, respectively, which are close to each other.
Figure~\ref{fig:Sbi_SprojplusSk}(b) shows the comparison between $\Delta S_{\mathrm{bi}}$ and $\Delta S_{\mathrm{proj}} + \Delta S_k$, where the latter is almost the same as $\Delta S_{\mathrm{bi},{\bf k}={\bf 0}} + \Delta S_k$ because $\Delta S_{\mathrm{proj}} \simeq \Delta S_{\mathrm{bi},{\bf k}={\bf 0}}$.
These two entanglement dynamics show a strong similarity, with a cosine similarity of $0.9871 < \cos \theta < 0.9882$ for $700\hbar /t_{\mathrm{e}} \leq \tau_{\mathrm{max}} \leq 1000\hbar /t_{\mathrm{e}}$.
Figure~\ref{fig:Sbi_SprojplusSk}(c) shows the amplitude spectra.
As seen by comparing Figs.~\ref{fig:Sbi_Sproj}(b) with \ref{fig:Sbi_SprojplusSk}(c), the difference between $|\Delta S_{\mathrm{bi}}(\Omega)|$ and $|\Delta S_{\mathrm{proj}}(\Omega)|$ around $\Omega = \Omega_{\mathrm{C}}$ originates from $\Delta S_k$.

Because $\Delta S_{\mathrm{bi}} (\tau) \simeq \Delta S_{\mathrm{proj}}(\tau) + \Delta S_k(\tau)$ holds, the entanglement dynamics obtained by DMS approximately corresponds to $\Delta S_{\mathrm{bi}}(\tau) - \Delta S_k(\tau)$.
Combining this with independent measurements of $\Delta S_k(\tau)$ allows for the full reconstruction of the biexciton entanglement dynamics.
In this sense, DMS is an effective method for investigating the entanglement dynamics of the biexciton.

Finally, we propose two experiments related to our simulation.
First, we consider how to measure $\Delta S_k(\tau)$.
To obtain $\Delta S_k(\tau)$ experimentally, both a spatial resolution and a time resolution are required.
The required spatial resolution is smaller than the radius of the biexciton (e.g., 1.5 nm for the CuCl crystal~\cite{CuCl_biradius}), which is much shorter than the optical wavelength. Therefore, obtaining information about $\Delta S_k(\tau)$ by standard optics using propagating light is challenging.
One possible experiment to measure $\Delta S_k(\tau)$ is to use near-field optics.
More specifically, dual-probe scanning near-field optical microscopy (Dual-SNOM) is a promising method.
After excitation by a propagating pump pulse with angular frequency $\omega_{\mathrm{pu}}$ at time $t=0$, a near-field probe pulse with angular frequency $\omega_{\mathrm{pr}}$ is focused and applied to the position ${\bf r}_{\mathrm{pr}}$ at time $t=\tau$. The resulting near-field generated at position ${\bf r}_{\mathrm{FWM}}$ with angular frequency $2\omega_{\mathrm{pu}}-\omega_{\mathrm{pr}}$ is measured.
By repeating the measurement with different $\Delta {\bf r} \coloneq {\bf r}_{\mathrm{FWM}}-{\bf r}_{\mathrm{pr}}$ and $\tau$, the spatially-resolved intensity $I(\Delta {\bf r},\tau)$ is obtained.
The Fourier transform of $I(\Delta {\bf r},\tau)$ with respect to $\Delta {\bf r}$, $I({\bf k},\tau)$, may be expressed as $I({\bf k},\tau) = F({\bf k})W_{{\bf k}}(\tau)$ where $F({\bf k})$ is an instrumental function reflecting the spatial resolution of the experimental system.
If the spatial resolution is sufficient to approximate $F({\bf k})$ as a constant, we can obtain $W_{{\bf k}}(\tau)$ and thus $S_k(\tau)$.

Second, we consider DMS using wurtzite semiconductors such as ZnO and GaN.
In these materials, there are several biexciton modes denoted $AA$, $AB$, and $BB$, where $A$ ($B$) denotes an exciton with a hole in the $A$- ($B$-) valence band~\cite{ZnO1,ZnO2,GaN}.
By simultaneously exciting several biexciton modes, quantum interference between different modes will occur, leading to oscillations in the entanglement dynamics of the biexciton, similar to those observed in the numerical simulation in this section.
DMS provides direct access to this oscillatory entanglement dynamics.

%% file: DMS_Summary.tex
\section{\label{SecV}Summary}
In this paper, we formulated DMS for both thick and thin samples.
DMS detects a pair of one-photon-allowed excitons forming a biexciton, whose wavevectors are the same as those of the probe light and the FWM light generated in the sample.
By performing numerical simulations of DMS using a model of a two-dimensional electron-hole system, we investigated the relation between the entanglement dynamics obtained by DMS and the biexciton's entanglement dynamics.
The dynamics of the EE obtained by DMS under normal incidence of light pulses, $\Delta S_{\mathrm{DMS}}(\tau)$ ($\simeq \Delta S_{\mathrm{proj}}(\tau)$), reflects the EE of $\ket{\psi_{\mathrm{bi},{\bf k}={\bf 0}}(\tau)}$, that is, $\Delta S_{\mathrm{proj}}(\tau) \simeq \Delta S_{\mathrm{bi},{\bf k} = {\bf 0}}(\tau)$.
In addition, we revealed the relation $\Delta S_{\mathrm{proj}}(\tau) \simeq \Delta S_{\mathrm{bi}}(\tau) - \Delta S_k (\tau)$, indicating that the entanglement obtained by DMS corresponds to the biexciton entanglement with the component associated with the exciton wavevectors removed.
In this sense, DMS is a useful method for investigating the entanglement dynamics of the biexciton.
We also propose near-field optical experiments to measure $\Delta S_k (\tau)$, which would enable us to reconstruct the full entanglement dynamics when combined with the DMS experiments.

\begin{acknowledgments}
The authors thank G. Oohata for fruitful discussions.
This work was supported by JSPS KAKENHI Grant Numbers JP23KJ0109 and JP24K00563. HM acknowledges support from CSIS, Tohoku University.
\end{acknowledgments}

%% file: DMS_AppendixA.tex
\section{\label{AppendixA}Derivation of ${\bf E}_{\mathrm{FWM}}$ in Sec.~\ref{SecIIA} from Maxwell's equations}
In this appendix, we derive Eq.~(\ref{eq:E_FWM}) in Sec.~\ref{SecIIA}.
Maxwell's equations in the Coulomb gauge are written as
\begin{eqnarray}
  &&\nabla^2 \phi({\bf r},\omega) = -\frac{\rho({\bf r},\omega)}{\varepsilon_0},  \label{eq:Maxwell_phi_thick}\\
  &&  \left( \nabla^2 + \frac{\omega^2}{c^2} \right) {\bf A}({\bf r},\omega) =
  -\mu_0 {\bf J}({\bf r},\omega) - \frac{i\omega}{c^2} \nabla \phi({\bf r},\omega),
  \label{eq:Maxwell_A_thick}
\end{eqnarray}
where $\phi$ (${\bf A}$) is the scalar (vector) potential, and $\rho$ (${\bf J}$) is the polarization charge (current) density.
The response of the magnetization is ignored.
By decomposing Eq.~(\ref{eq:Maxwell_A_thick}) into longitudinal and transverse components, we can obtain
\begin{eqnarray}
  &&{\bf J}_{\mathrm{L}}({\bf r},\omega) = -i\varepsilon_0 \omega \nabla \phi({\bf r},\omega), \label{eq:Maxwell_J_L}\\
  &&\left( \nabla^2 + \frac{\omega^2}{c^2} \right) {\bf A}({\bf r},\omega) = -\mu_0 {\bf J}_{\mathrm{T}}({\bf r},\omega),
  \label{eq:Maxwell_J_T} 
\end{eqnarray}	
where ${\bf J}_{\mathrm{L}}$ (${\bf J}_{\mathrm{T}}$) is the longitudinal (transverse) component of the current density.
Considering the charge conservation law $-i\omega\rho+\divergence {\bf J}_{\mathrm{L}} = 0$, Eq.~(\ref{eq:Maxwell_J_L}) corresponds to Eq.~(\ref{eq:Maxwell_phi_thick}).
Hereafter, we focus on Eq.~(\ref{eq:Maxwell_J_T}) because far fields are composed of only the transverse components.
The transverse current density can be expressed as
\begin{equation}
  {\bf J}_{\mathrm{T}}({\bf r},\omega) =\varepsilon_0\omega^2\chi(\omega){\bf A}({\bf r},\omega) + {\bf J}_{\mathrm{NL},\mathrm{T}}({\bf r},\omega).
  \label{eq:linear_susceptibility}
\end{equation}
Here, the first term on the right-hand side represents a linear response to the light, and $\chi(\omega)$ is the linear susceptibility related to the refractive index as $n(\omega) = \sqrt{1+\chi(\omega)}$.
The second term on the right-hand side represents a nonlinear response to the light.
By substituting Eq.~(\ref{eq:linear_susceptibility}) into Eq.~(\ref{eq:Maxwell_J_T}), we can obtain
\begin{equation}
  \left(\nabla^2+\frac{n(\omega)^2\omega^2}{c^2}\right) {\bf A}({\bf r},\omega) =
  -\mu_0 {\bf J}_{\mathrm{NL},\mathrm{T}}({\bf r},\omega).
  \label{eq:Helmholtz}
\end{equation}
This equation shows that the transmitted FWM light is generated by the nonlinear transverse current.

Here, the nonlinear current density ${\bf J}_{\mathrm{NL}}$ is assumed to be 
\begin{equation}
  {\bf J}_{\mathrm{NL}}({\bf r},\omega) = \sum_{{\bf k}}{\bf J}_{\mathrm{NL},{\bf k}}(z,\omega)\exp(i{\bf k} \cdot {\bf r})\theta(z)\theta(L_z-z),
\end{equation}
where $\theta(z)$ is a step function.
The corresponding nonlinear transverse current density ${\bf J}_{\mathrm{NL,T}}$ can be expressed as
\begin{equation}
  {\bf J}_{\mathrm{NL},\mathrm{T}}({\bf r},\omega) = \sum_{{\bf k}}{\bf J}_{\mathrm{NL},\mathrm{T},{\bf k}}(z,\omega)\exp(i{\bf k} \cdot {\bf r}) \theta(z) \theta(L_z-z),
\end{equation}
where the transverse component of ${\bf J}_{\mathrm{NL},{\bf k}}(\omega)$ is denoted by ${\bf J}_{\mathrm{NL},\mathrm{T},{\bf k}}(z,\omega)$. To properly treat the boundary conditions, ${\bf J}_{\mathrm{NL},\mathrm{T},{\bf k}}$ is decomposed into the bulk and boundary terms:
\begin{align}
  {\bf J}_{\mathrm{NL},\mathrm{T},{\bf k}}(z,\omega) &= {\bf J}_{\mathrm{NL},\mathrm{T},{\bf k},\mathrm{bulk}}(\omega) + {\bf J}_{\mathrm{NL},\mathrm{T},{\bf k},\mathrm{s},0}(z,\omega) \nonumber \\
  &\quad + {\bf J}_{\mathrm{NL},\mathrm{T},{\bf k},\mathrm{s},L_{z}}(z,\omega),
\end{align}
where each term on the right-hand side satisfies
\begin{align}
  &{\bf J}_{\mathrm{NL},\mathrm{T},{\bf k},\mathrm{bulk}}(\omega) = {\bf J}_{\mathrm{NL},{\bf k}} - ({\bf e}_{{\bf k}} \cdot {\bf J}_{\mathrm{NL},{\bf k}}){\bf e}_{{\bf k}}, \\
  &|{\bf J}_{\mathrm{NL},\mathrm{T},{\bf k},\mathrm{s},0}(z,\omega)| \propto e^{-k_{\parallel}z}, \\
  &|{\bf J}_{\mathrm{NL},\mathrm{T},{\bf k},\mathrm{s},L_{z}}(z,\omega)| \propto e^{-k_{\parallel}(L_{z}-z)}.
\end{align}
${\bf J}_{\mathrm{NL},\mathrm{T},{\bf k},\mathrm{s},0}(z,\omega)$ and ${\bf J}_{\mathrm{NL},\mathrm{T},{\bf k},\mathrm{s},L_{z}}(z,\omega)$ are the transverse current terms localized near the surfaces, which originate from $\divergence {\bf J}_{\mathrm{NL,T}} = 0$ at the surfaces ($J^z_{\mathrm{NL,T}}(z=0) = J^z_{\mathrm{NL,T}}(z=L_{z}) = 0$).
Note that ${\bf J}_{\mathrm{NL},\mathrm{T},{\bf k},\mathrm{s},0} = {\bf J}_{\mathrm{NL},\mathrm{T},{\bf k},\mathrm{s},L_{z}} = 0$ for $k_{\parallel} = 0$.

The vector potential of the light emitted from ${\bf J}_{\mathrm{NL,T}}$, $\Delta {\bf A}$, can be expressed as
\begin{equation}
  \Delta{\bf A}({\bf r},\omega) = \sum_{\bf k_{\parallel}}{\bf a}_{\bf k_{\parallel}}(z,\omega)e^{i{\bf k}_{\parallel}\cdot {\bf r}_{\parallel}}.
\end{equation}
Then Eq.~(\ref{eq:Helmholtz}) becomes
\begin{equation}
  \left(\frac{\partial^2 }{\partial z^2} + k_0^2 \right){\bf a}_{{\bf k}_{\parallel}}(z,\omega) = -\mu_0 \sum_{k_z} {\bf J}_{\mathrm{NL},\mathrm{T},{\bf k}}(z,\omega)e^{ik_z z}\theta(z)\theta(L_z-z).
\end{equation}
To solve this equation, we introduce the Green's function $G(z-z')$:
\begin{equation}
  \left(\frac{\partial^2 }{\partial z^2} + k_0^2 \right)G(z-z') = -\delta(z-z').
\end{equation}
The solution of this differential equation can be expressed as $G(z-z') = g_{+} e^{ik_0|z-z'|} + g_{-}e^{-ik_0|z-z'|}$ where $g_+ - g_- = i/(2k_0)$.
Because we consider the emission of light from the source, we choose $g_+ = i/(2k_0)$ and $g_- = 0$:
\begin{equation}
  G(z-z') = \frac{i}{2k_0} e^{ik_0|z-z'|}.
\end{equation}
Using the Green's function, ${\bf a}_{\bf k_{\parallel}}(z,\omega)$ can be obtained as
\begin{equation}
  {\bf a}_{{\bf k}_{\parallel}}(z,\omega) = \mu_0 \int_{0}^{L_{z}} \mathrm{d} z' \ G(z-z')\sum_{k_z}{\bf J}_{\mathrm{NL},\mathrm{T},{\bf k}}(z',\omega) e^{ik_z z'}.
  \label{eq:J_to_a}
\end{equation}
This solution satisfies the Coulomb-gauge condition: $\divergence \Delta {\bf A} = 0$.

We assume that the emitted light from ${\bf J}_{\mathrm{NL},\mathrm{T},{\bf k},\mathrm{bulk}}(\omega)$ is the dominant contribution to $\Delta {\bf A}|_{z=L_z}$, and we replace ${\bf J}_{\mathrm{NL},\mathrm{T},{\bf k}}$ with ${\bf J}_{\mathrm{NL},\mathrm{T},{\bf k},\mathrm{bulk}}$ at Eq.~(\ref{eq:J_to_a}).
Then ${\bf a}_{{\bf k}}(z,\omega)$ can be expressed as
\begin{align}
  &{\bf a}_{{\bf k}_{\parallel}}(z,\omega) \equiv {\bf a}_{+,{\bf k}_{\parallel}}(z,\omega) + {\bf a}_{-,{\bf k}_{\parallel}}(z,\omega), \nonumber \\
  &{\bf a}_{+,{\bf k}_{\parallel}}(z,\omega) = \mu_0\int^z_0 \mathrm{d}z' \ G(z-z') \sum_{k_z}{\bf J}_{\mathrm{NL},\mathrm{T},{\bf k},\mathrm{bulk}}(\omega) e^{ik_z z'}, \nonumber \\
  &{\bf a}_{-,{\bf k}_{\parallel}}(z,\omega) = \mu_0\int^{L_z}_z \mathrm{d}z' \ G(z-z')\sum_{k_z} {\bf J}_{\mathrm{NL},\mathrm{T},{\bf k},\mathrm{bulk}}(\omega) e^{ik_z z'}.
\end{align}
If multiple reflections are ignored, only ${\bf a}_{+,{\bf k}}$ contributes to the light transmitted from the sample at $z=L_z$.
In this paper, the incident light comes from $z<0$ and the transmitted light at $z>L_z$ is detected.
Therefore, we only consider ${\bf a}_{+,{\bf k}}(z,\omega)$.
The vector potential of light propagating toward $z>0$ in the sample, $\Delta {\bf A}_+$, becomes
\begin{align}
  \Delta{\bf A}_{+}({\bf r},\omega)
  &= \sum_{{\bf k}_{\parallel}}{\bf a}_{+,{\bf k}_{\parallel}}(z,\omega)e^{i{\bf k}_{\parallel} \cdot {\bf r}_{\parallel}} \nonumber \\
  &= \sum_{{\bf k}}\frac{i\mu_0 {\bf J}_{\mathrm{NL},\mathrm{T},{\bf k},\mathrm{bulk}}}{2k_0}
    z \sinc\left(\frac{\Delta k}{2}z \right) e^{i{\bf k}\cdot {\bf r} + i\frac{\Delta k}{2}z}.
    \label{eq:Appendix_A+}
\end{align}
The transverse electric field $\Delta {\bf E}_{\mathrm{T,tr}}({\bf r},\omega)$ of the transmitted light ($z>L_z$) can be obtained by decomposing the light at $z=L_z-0$ into $s$- and $p$-polarized components and multiplying the amplitude transmittance $t_{\mathrm{sv},{\bf k}_0}^\alpha$ ($\alpha=s,p$):
\begin{align}
  \Delta E_{\mathrm{T,tr}}^\alpha({\bf r},\omega)
  &= \sum_{{\bf k}_{\parallel}}i\omega t_{\mathrm{sv},{\bf k}_0}^\alpha a_{+,{\bf k}_{\parallel}}(z=L_z,\omega)e^{i{\bf k}_{\parallel} \cdot {\bf r}_{\parallel}} e^{ik_{\mathrm{tr}}(z-L_{z})} \nonumber \\
  &\equiv \sum_{\bf k}t_{\mathrm{sv},{\bf k}_0}^\alpha f_{\bf k}(\omega) J_{\mathrm{NL},\mathrm{T},{\bf k},\mathrm{bulk}}^\alpha e^{i{\bf k}_{\mathrm{tr}} \cdot {\bf r}},
  \label{eq:DeltaE_final}
\end{align}
where $f_{\bf k}(\omega)$ is defined as Eq.~(\ref{eq:f_k_def}).
By using Eq.~(\ref{eq:DeltaE_final}) and considering FWM, Eq.~(\ref{eq:E_FWM}) in the main text can be obtained.

%% file: DMS_AppendixB.tex
\section{\label{AppendixB}The effects of multiple reflections}
In this appendix, we describe the effects of multiple reflections neglected in Sec.~\ref{SecII} and Appendix~\ref{AppendixA}.
When we incorporate the effects of multiple reflections, Equation~(\ref{eq:Appendix_A+}) is replaced with
  \begin{align}
    \Delta {\bf A}_{+} ({\bf r},\omega)
    &\equiv \sum_{{\bf k}_{\parallel}}
    {\bf a}_{+,{\bf k}_{\parallel}}'(z,\omega)e^{i {\bf k}_{\parallel} \cdot {\bf r}_{\parallel}} , \\
     {\bf a}_{+,{{\bf k}_{\parallel}}}' (z,\omega) 
    &\equiv {\bf a}_{+,{{\bf k}_{\parallel}}} (z,\omega) \nonumber \\
    &\hspace{0.5em}+ \sum_{m=1}^\infty {\bf a}_{+,{{\bf k}_{\parallel}},m} (z,\omega) + \sum_{m=1}^\infty {\bf a}_{-,{{\bf k}_{\parallel}},m} (z,\omega),  \\
     a_{+,{{\bf k}_{\parallel}},m}^\alpha (z,\omega) 
    &= \mu_0 (r_{\mathrm{sv},{\bf k}_0}^\alpha)^{2m} 
    \sum_{k_z}J^\alpha_{\mathrm{NL},\mathrm{T},{\bf k},\mathrm{bulk}}(\omega)  \nonumber \\
    &\hspace{1em} \times \int^{L_z}_0 \mathrm{d}z' \ G(z+2mL_z-z')  e^{ik_z z'}, \\
    a_{-,{{\bf k}_{\parallel}},m}^\alpha (z,\omega) 
    &= \mu_0 (r_{\mathrm{sv},{\bf k}_0}^\alpha)^{2m-1} 
    \sum_{k_z}J^\alpha_{\mathrm{NL},\mathrm{T},{\bf k},\mathrm{bulk}}(\omega) \nonumber \\
    &\hspace{1em}\times \int^{L_z}_0 \mathrm{d}z' \ G(-z-2mL_z-z')  e^{ik_z z'},
  \end{align}
where $r_{\mathrm{sv},{\bf k}_0}^\alpha$ is an amplitude reflectance of $\alpha$-polarized light with the wavevector ${\bf k}_0$.
${\bf a}_{+,{\bf k},m}(z,\omega)$ arises from ${\bf a}_{+,{\bf k}}(L_z,\omega)$ after $2m$ reflections at the boundaries of the sample.
Similarly, ${\bf a}_{-,{\bf k},m}(z,\omega)$ arises from ${\bf a}_{-,{\bf k}}(0,\omega)$ after $(2m-1)$ reflections at the boundaries of the sample.
The transverse electric field of the transmitted light can be obtained by multiplying the amplitude transmittance $t_{\mathrm{sv},{\bf k}_{0}}^\alpha$:
\begin{align}
  &\Delta E_{\mathrm{T},\mathrm{tr}}^\alpha({\bf r},\omega) =
  \sum_{\bf k}i\omega t_{\mathrm{sv},{\bf k}_0}^\alpha {a'}_{+,{\bf k}}^{\alpha}(z=L_{z},\omega)e^{i{\bf k}_{\parallel} \cdot {\bf r}_{\parallel}}e^{ik_{\mathrm{tr}}(z-L_{z})} \nonumber \\
  &= \sum_{\bf k} \frac{t_{\mathrm{sv},{\bf k}_0}^\alpha}{1-(r_{\mathrm{sv},{\bf k}_0}^\alpha)^2e^{2ik_0L_z}}f_{{\bf k}}(\omega)J_{\mathrm{NL,T},{\bf k},\mathrm{bulk}}^\alpha (\omega)e^{i{\bf k}_{\mathrm{tr}}\cdot {\bf r} } \nonumber \\
  &+ \sum_{\bf k} \frac{t_{\mathrm{sv},{\bf k}_0}^\alpha r_{\mathrm{sv},{\bf k}_0}^\alpha}{1-(r_{\mathrm{sv},{\bf k}_0}^\alpha)^2e^{2ik_0L_z}}f_{{\bf k},-}(\omega) J_{\mathrm{NL,T},{\bf k},\mathrm{bulk}}^\alpha (\omega) e^{i{\bf k}_{\mathrm{tr}}\cdot {\bf r} },
  \label{eq:DeltaE_tr_multiref}
\end{align}
where
\begin{align}
  &f_{{\bf k},-}(\omega)=
  -\frac{\mu_0 \omega}{2k_0} L_z \sinc\left(\frac{\Delta k_-}{2}L_z \right)
  e^{ i(k_0-k_{\mathrm{tr}} +\Delta k_-/2)L_z}, \\
  &\Delta k_- \coloneq k_0 + k_z .
\end{align}

The vector potential of the probe pulse in the sample becomes
\begin{align}
  A_{\mathrm{pr}}^\alpha({\bf r},\omega) =
  &\sum_{{\bf k}_{\parallel}} 
  \frac{t_{\mathrm{vs},{\bf k}_0}^\alpha A_{\mathrm{in},{\bf k}_{\parallel}}^{\alpha}(\omega) e^{i {\bf k}_{\parallel} \cdot {\bf r}_{\parallel} } }{1-(r_{\mathrm{sv},{\bf k}_0}^{\alpha})^2e^{2ik_0L_{z}}}  \nonumber \\
& \hspace{1em}\times \left[
    e^{ik_0z} + r_{\mathrm{sv},{\bf k}_0}^\alpha e^{ik_0(2L_z-z)}
  \right].
\end{align}
The corresponding Fourier component ${\bf A}_{{\bf k}}$ is
\begin{align}
  A^\alpha_{{\bf k}}(\omega) 
  &=
  \frac{A_{\mathrm{in},{\bf k}_{\parallel}}^{\alpha}(\omega)t_{\mathrm{vs},{\bf k}_0}^\alpha}{1-(r_{\mathrm{sv},{\bf k}_0}^\alpha)^2 e^{2ik_0L_{z}} }
  \left[ \frac{e^{i(k_0-k_z)L_z}-1}{i(k_0-k_z)L_z}  \right. \nonumber \\
    & \hspace{1em}\left. - r_{\mathrm{sv},{\bf k}_0}^\alpha e^{2ik_0L_z} \frac{e^{-i(k_0+k_z)L_z}-1}{i(k_0+k_z)L_z}
    \right].
\end{align}

The wavevector of the pump pulse is denoted by ${\bf k}_{\mathrm{pu},+}$ (${\bf k}_{\mathrm{pu},-}$), whose $z$-component is positive (negative).
${\bf k}_{\mathrm{pu},+}$ and ${\bf k}_{\mathrm{pu},-}$ are related as
\begin{align}
  &  {\bf k}_{\mathrm{pu},+,z} = -{\bf k}_{\mathrm{pu},-,z} > 0,  \\
  & {\bf k}_{\mathrm{pu},+,j} = {\bf k}_{\mathrm{pu},-,j} \ \ \ (j=x,y).
\end{align}
$\ket{\psi_{\mathrm{pu}}(\tau)}$ can be expressed with
\begin{align}
  \ket{\psi_{\mathrm{pu}}(\tau)} = &\sqrt{W_0}\exp \left( -i\frac{\epsilon_0}{\hbar}\tau \right) \ket{0}+ \sqrt{W_{\mathrm{bi,+}}}\ket{\psi_{\mathrm{bi,+}}(\tau)} \nonumber \\
  & +  \sqrt{W_{\mathrm{bi,-}}}\ket{\psi_{\mathrm{bi,-}}(\tau)},
\end{align}
where $\ket{\psi_{\mathrm{bi,+}}}$ ($\ket{\psi_{\mathrm{bi,-}}}$) is the biexciton state with wavevector ${\bf k}_{\mathrm{pu},+}$ (${\bf k}_{\mathrm{pu},-}$), created via two-photon absorption of the pump pulse, and $W_{\mathrm{bi},+}$ ($W_{\mathrm{bi},-}$) is the corresponding weight.
$\ket{\psi_{\mathrm{bi,\pm}}}$ satisfies 
\begin{align}
  \hat{T}_{{\bf R}} \ket{\psi_{\mathrm{bi},\pm}(\tau)} \simeq \exp(-2i{\bf k}_{\mathrm{pu},\pm}\cdot {\bf R})\ket{\psi_{\mathrm{bi}}(\tau)}.
\end{align}
The transverse current density related to FWM becomes
\begin{align}
  &{\bf J}_{\mathrm{T},\mathrm{FWM},{\bf k}} (\omega,\tau) \equiv {\bf J}_{\mathrm{T},\mathrm{FWM},{\bf k},+} (\omega,\tau) + {\bf J}_{\mathrm{T},\mathrm{FWM},{\bf k},-} (\omega,\tau), \label{eq:J_T_FWM_multiref}\\
  &J_{\mathrm{T},\mathrm{FWM},{\bf k},\pm}^\alpha (\omega,\tau)
  =
  V\sqrt{W_0 W_{\mathrm{bi},\pm}}
  \exp\left[i \left(\omega + \frac{\epsilon_0}{\hbar} \right) \tau \right] \nonumber \\
  &\times \sum_{\beta = s,p}
  \braket{\phi_{{\bf k},\pm}^{\alpha\beta}|\psi_{\mathrm{bi},\pm}(\tau)} 
  \left[
    A^\beta_{2{\bf k}_{\mathrm{pu},\pm}-{\bf k}}
    \left(  \frac{\epsilon_{\mathrm{bi}}}{\hbar}-\omega, 0   \right)
    \right]^* ,
    \label{eq:J_T_FWM_pm_multiref}
\end{align}
where $\bra{\phi_{{\bf k},\pm}^{\alpha\beta}}$ is defined as
\begin{align}
  \bra{\phi_{{\bf k},\pm}^{\alpha \beta}} &\coloneq \bra{0}\hat{J}^\beta_{\mathrm{T},2{\bf k}_{\mathrm{pu},\pm}-{\bf k}}  \frac{1}{\hbar \omega - \epsilon_{\mathrm{bi}}+(\hat{H}_0-\epsilon_0)+i\eta}\hat{J}^\alpha_{\mathrm{T},{\bf k}}  \nonumber \\
  &\hspace{1em}-\bra{0} \hat{J}^\alpha_{\mathrm{T},{\bf k}}  \frac{1}{\hbar \omega - (\hat{H}_0-\epsilon_0)+i\eta} \hat{J}^\beta_{\mathrm{T},2{\bf k}_{\mathrm{pu},\pm}-{\bf k}}.
\end{align}

By using Eqs.~(\ref{eq:DeltaE_tr_multiref}) and (\ref{eq:J_T_FWM_multiref}), FWM-electric field can be written as 
\begin{align}
  &\hspace{-1em}{\bf E}_{\mathrm{FWM}}({\bf r},\omega) = \sum_{{\bf k}_{\parallel}} {\bf E}_{\mathrm{FWM},{\bf k}_{\parallel}} e^{i{\bf k}_{\mathrm{tr}} \cdot {\bf r}} (\omega,\tau), \\
  &\hspace{-1em}{\bf E}_{\mathrm{FWM},{\bf k}_{\parallel}} (\omega,\tau)  \equiv   {\bf E}_{\mathrm{FWM},{\bf k}_{\parallel},+} (\omega,\tau)  + {\bf E}_{\mathrm{FWM},{\bf k}_{\parallel},-} (\omega,\tau),  \\
  &\hspace{-1em}E_{\mathrm{FWM},{\bf k}_{\parallel},\pm}^\alpha(\omega,\tau)
  = \sum_{k_z} \left[\frac{t_{\mathrm{sv},{\bf k}_0}^\alpha f_{{\bf k}}(\omega)J_{\mathrm{T},\mathrm{FWM},{\bf k},\pm}^\alpha (\omega,\tau)}{1-(r_{\mathrm{sv},{\bf k}_0}^\alpha)^2e^{2ik_0L_{z}}}  \right. \nonumber \\
  &\hspace{5em}\left. + \frac{t_{\mathrm{sv},{\bf k}_0}^\alpha r_{\mathrm{sv},{\bf k}_0}^\alpha f_{{\bf k},-}(\omega)J_{\mathrm{T},\mathrm{FWM},{\bf k},\pm}^\alpha (\omega,\tau)}{1-(r_{\mathrm{sv},{\bf k}_0}^\alpha)^2e^{2ik_0L_{z}}} \right].
  \label{eq:E_FWM_multiref}
\end{align}
By combining Eqs.~(\ref{eq:J_T_FWM_pm_multiref}) and (\ref{eq:E_FWM_multiref}), the relation between the FWM-electric field ${\bf E}_{\mathrm{FWM}}$ and the biexciton state $\ket{\psi_{\mathrm{bi}, \pm}}$ can be obtained.

If the phase-matching condition described below is satisfied, DMS can be carried out.
We denote the central wavevector of the probe pulse in the sample as ${\bf k}_{\mathrm{pr},+}$ (${\bf k}_{\mathrm{pr},-}$) whose $z$ component is positive (negative).
The phase-matching condition of FWM including reflected light is described as
\begin{equation}
  s_3k_0 =  2k_{\mathrm{pu},s_1,z}-k_{\mathrm{pr},s_2,z} \ \ \ (s_i = +, -),
\end{equation}
where $s_1 = +$ ($-$) if the $z$ component of the wavevector of the used pump pulse is positive (negative), $s_2 = +$ ($-$) if the $z$ component of the wavevector of the used probe pulse is positive (negative), and $s_3 = +$ ($-$) if the number of the reflection of the generated FWM light is even (odd).
We assume that the process corresponding to a specific $(s_1,s_2,s_3)$ contributes predominantly to the FWM light at $z>L_z$.
Then the polarized intensity of the FWM signal takes the form
\begin{align}
  &|{\bf e}_{1}\cdot {\bf E}_{\mathrm{FWM},{\bf k}_{\parallel}}(\omega,\tau)|^2 \nonumber\\
  &\simeq \mathcal{N}_{s_1,s_2,s_3} \left |
  \sum_{\alpha,\beta=s,p} (e_{1}^\alpha e_{2}^\beta)^* 
  \psi_{{\bf k}_{\parallel},s_1,s_2,s_3}^{\alpha \beta}(\omega,\tau)
  \right |^2,
  \label{eq:polarized-intensity_multiref}
\end{align}
where $\psi^{\alpha \beta}_{{\bf k}_{\parallel},s_1,s_2,s_3}(\omega,\tau)$ is defined as

\begin{align}
  &\psi^{\alpha \beta}_{{\bf k}_{\parallel},s_1,s_2,s_3}(\omega,\tau)\nonumber \\
  &= \frac{p_{s_3,{\bf k}_{s_3}}^\alpha  \left(q_{s_2,{\bf k}_{\mathrm{pr},s_2}}^\beta \right)^*   \braket{\phi_{{\bf k}_{s_3}^*,s_1}^{\alpha\beta}(\omega)|\psi_{\mathrm{bi},s_1}(\tau)}}{\sqrt{\sum_{\alpha'\beta'}
      \left| p_{s_3,{\bf k}_{s_3}}^{\alpha'} \right|^2
      \left| q_{s_2,{\bf k}_{\mathrm{pr},s_2}}^{\beta'}\right|^2
      \left |
      \braket{\phi_{{\bf k}_{s_3}^*,s_1}^{\alpha'\beta'}|\psi_{\mathrm{bi},s_1}(\tau)}
      \right |^2 }}.
\end{align}
Here, ${\bf k}_{s_3}$ and ${\bf k}_{s_3}^*$ are defined as ${\bf k}_{s_3} \coloneq {\bf k}_{\parallel} + s_3 k_0{\bf e}_z$ and ${\bf k}_{s_3}^* = {\bf k}_{\parallel} + k_z^* {\bf e}_z$, respectively, where 
$k_z^* = \argmin_{k_z}|k_z - s_3 k_0|$.
$p_{s_3,{\bf k}_{s_3}}^\alpha$ and $q_{s_2,{\bf k}_{\mathrm{pr},s_2}}^\beta$ are defined as
\begin{align}
  & p_{s_3,{\bf k}_{s_3}}^\alpha \coloneq \frac{t_{\mathrm{sv},{\bf k}_{s_3}}^\alpha \left(r_{\mathrm{sv},{\bf k}_{s_3}}^\alpha\right)^{(1-s_3)/2}}{1-\left(r_{\mathrm{sv},{\bf k}_{s_3}}^\alpha\right)^2 e^{2ik_0 L_z}},   \\
  & q_{s_2,{\bf k}_{\mathrm{pr},s_2}}^\beta \coloneq \frac{t_{\mathrm{vs},{\bf k}_{\mathrm{pr,tr}}}^\beta \left(r_{\mathrm{sv},{\bf k}_{\mathrm{pr},s_2}}^\beta \right)^{(1-s_2)/2}}{1-\left(r_{\mathrm{sv},{\bf k}_{\mathrm{pr},s_2}}^\beta \right) ^2e^{2ik_{\mathrm{pr},+,z}L_z}}.
\end{align}
By performing the tomography as described in Sec.~\ref{SecIIB}, $\psi^{\alpha \beta}_{{\bf k}_{\parallel},s_1,s_2,s_3}(\omega,\tau)$ can be determined experimentally, except for a polarization-independent phase factor.

%% file: DMS_AppendixC.tex
\section{\label{AppendixC}Derivation of ${\bf E}_{\mathrm{FWM}}$ in Sec.~\ref{SecIII} from Maxwell's equations}
In this appendix, we derive Eq.~(\ref{eq:E_thin}) in Sec.~\ref{SecIII}.
The charge density derived from the charge conservation law using Eq.~(\ref{eq:J_thin}) is given by
\begin{align}
  &\rho({\bf r},\omega) = \sigma({\bf r}_{\parallel},\omega)\delta(z) = \sum_{{\bf k}_{\parallel}}\sigma_{{\bf k}_{\parallel}}(\omega)e^{i{\bf k}_{\parallel} \cdot {\bf r}_{\parallel}} \delta(z), \nonumber \\
  & \sigma_{{\bf k}_{\parallel}}(\omega) = \frac{{\bf k}_{\parallel} \cdot {\bf j}_{{\bf k}_{\parallel}}}{\omega}  \ \ \ (\sigma_{{\bf k}_{\parallel} = {\bf 0}}= 0 ).
  \label{eq:rho_thin}
\end{align}
Fourier expansions of $\phi$ and ${\bf A}$ with respect to $x$ and $y$ are
\begin{align}
  &\phi({\bf r},\omega) = \sum_{{\bf k}_{\parallel}}\phi_{{\bf k}_{\parallel}}(z,\omega)e^{i{\bf k}_{\parallel}\cdot {\bf r}_{\parallel}},  \label{eq:FT_phi_thin} \\
  & {\bf A}({\bf r},\omega) = \sum_{{\bf k}_{\parallel}}{\bf A}_{{\bf k}_{\parallel}}(z,\omega)e^{i{\bf k}_{\parallel} \cdot {\bf r}_{\parallel}}.
  \label{eq:FT_A_thin}
\end{align}
By substituting Eqs.~(\ref{eq:rho_thin}) and (\ref{eq:FT_phi_thin}) into Eq.~(\ref{eq:Maxwell_phi_thick}), we can obtain
\begin{equation}
  \left( \frac{\partial^2}{\partial z^2} - k_{\parallel}^2  \right) \phi_{{\bf k}_{\parallel}}(z,\omega)
  = -\frac{{\bf k}_{\parallel} \cdot {\bf j}_{{\bf k}_{\parallel}}}{\varepsilon_0 \omega}\delta(z),
\end{equation}
where $k_{\parallel} = |{\bf k}_{\parallel}|$.
The solution satisfying $\phi_{{\bf k}_{\parallel}}(z=\pm \infty,\omega) = 0$ is given by
\begin{equation}
  \phi_{{\bf k}_{\parallel}}(z,\omega) = \frac{{\bf e}_{{\bf k}_{\parallel}} \cdot {\bf j}_{{\bf k}_{\parallel}}}{2\varepsilon_0 \omega} e^{-k_{\parallel}|z|},
  \label{eq:phi_thin}
\end{equation}
where ${\bf e}_{{\bf k}_{\parallel}} \coloneq {\bf k}_{\parallel}/|{\bf k}_{\parallel}|$.
By substituting Eqs.~(\ref{eq:J_thin}), (\ref{eq:FT_A_thin}), and (\ref{eq:phi_thin}) into Eq.~(\ref{eq:Maxwell_A_thick}), we can obtain
\begin{align}
  & \left( \frac{\partial^2}{\partial z^2} + k_{\mathrm{tr}}^2 \right)
  {\bf A}_{{\bf k}_{\parallel}}(z,\omega) =
  -\mu_0 {\bf j}_{{\bf k}_{\parallel}}(\omega)\delta(z) \nonumber \\
  &\hspace{4em}+ \frac{\mu_0 k_{\parallel}}{2} ({\bf e}_{{\bf k}_{\parallel}} \cdot {\bf j}_{{\bf k}_{\parallel}})
  \left[ {\bf e}_{{\bf k}_{\parallel}} + i \sgn (z){\bf e}_{z} \right] e^{-k_{\parallel}|z|}.
\end{align}
The corresponding Green's function $G_{\mathrm{thin}}(z-z')$ is given by
\begin{align}
  &\left( \frac{\partial^2}{\partial z^2} + k_{\mathrm{tr}}^2 \right) G_{\mathrm{thin}}(z-z') = -\delta(z-z'),  \\
  & G_{\mathrm{thin}}(z-z') = \frac{i}{2k_{\mathrm{tr}}} e^{i k_{\mathrm{tr}} |z-z'|}.
\end{align}
The vector potential of the emitted light from the sample, $\Delta {\bf A}_{{\bf k}_{\parallel}}$, can be expressed as
\begin{align}
  &\Delta {\bf A}_{{\bf k}_{\parallel}}(z,\omega) = \int^\infty_{-\infty}\mathrm{d}z' \
  G_{\mathrm{thin}}(z-z') \bigg \{
  \mu_0 {\bf j}_{{\bf k}_{\parallel}}(\omega)\delta(z')  \nonumber \\
  &\hspace{3em}
  - \frac{\mu_0 k_{\parallel}}{2} ({\bf e}_{{\bf k}_{\parallel}} \cdot {\bf j}_{{\bf k}_{\parallel}})
  [ {\bf e}_{{\bf k}_{\parallel}} + i \sgn (z'){\bf e}_{z} ] e^{-k_{\parallel}|z'|} \bigg \}.
\end{align}
By performing $z'$ integral and neglecting terms proportional to $e^{-k_{\parallel}|z|}$, we can obtain
\begin{align}
  &\Delta {\bf A}_{{\bf k}_{\parallel}}(z,\omega) \simeq \frac{i\mu_0}{2k_{\mathrm{tr}}} {\bf j}_{\mathrm{T},{\bf k}_{\mathrm{tr}}}(\omega)e^{ik_{\mathrm{tr}}z}\ \ \ (z \gg 1/|{\bf k}_{\parallel}|), \\ 
  &\Delta {\bf A}_{{\bf k}_{\parallel}={\bf 0}}(z,\omega) = \frac{i\mu_0}{2k_{\mathrm{tr}}} {\bf j}_{\mathrm{T},{\bf k}_{\mathrm{tr}}}(\omega)e^{ik_{\mathrm{tr}}z} \ \ \ (z>0),
\end{align}
where ${\bf j}_{\mathrm{T},{\bf k}_{\mathrm{tr}}}$ is defined by Eq.~(\ref{eq:j_T_def}).
Using $\Delta {\bf E}_{\mathrm{T},{\bf k}_{\parallel}} = i\omega \Delta {\bf A}_{{\bf k}_{\parallel}}$, the electric field of the generated light can be expressed by Eq.~(\ref{eq:E_thin}).
Note that the longitudinal component of the produced electric field ${\bf E}_{\mathrm{L}} = -\nabla \phi$ shows an exponential decay with respect to $|z|$ (${\bf E}_{\mathrm{L},{\bf k}_{\parallel}} \propto e^{-k_{\parallel}|z|}$), and it is negligible at $z \gg 1/|{\bf k}_{\parallel}|$.

%% file: DMS_AppendixD.tex
\section{\label{AppendixD}Derivation of the decomposition formula for $S_{\mathrm{bi}}$}

The reduced density matrix of the exciton A is given by
\begin{align}
  \hat{\rho}_{\mathrm{A}} &= \Tr_{\mathrm{B}} \left(\ket{\psi_{\mathrm{bi}}} \bra{\psi_{\mathrm{bi}}}\right) \nonumber \\
  &= \sum_{{\bf k}j_{\mathrm{A}}j_{\mathrm{A}}'} \left(
  \sum_{j_{\mathrm{B}}} \psi_{{\bf k}j_{\mathrm{A}}j_{\mathrm{B}}} \psi_{{\bf k}j_{\mathrm{A}}' j_{\mathrm{B}}}^*
  \right)
  \ket{{\bf k},j_{\mathrm{A}}} \bra{{\bf k},j_{\mathrm{A}}'}  \nonumber \\
  &= \sum_{{\bf k}}W_{{\bf k}}\Tr_{\mathrm{B}}\left(\ket{\psi_{\mathrm{bi},{\bf k}}} \bra{\psi_{\mathrm{bi},{\bf k}}}\right) \nonumber \\
  &\equiv \sum_{{\bf k}}W_{{\bf k}} \hat{\rho}_{\mathrm{A},{\bf k}}.
\end{align}
Here, we consider the matrix representation of $\hat{\rho}_{\mathrm{A}}$ in the basis $\{ \ket{{\bf k}, j_{\mathrm{A}}} \}$. 
The matrix $\hat{\rho}_{\mathrm{A}}$ takes a block-diagonal form, where each block corresponds to a ${\bf k}$ sector of size $j_{\mathrm{max}}\times j_{\mathrm{max}}$:
\begin{align}
  \hat{\rho}_{\mathrm{A}}
  &= \sum_{{\bf k}}W_{{\bf k}}\hat{\rho}_{\mathrm{A},{\bf k}} \nonumber \\
  &= \begin{bmatrix}
    W_{{\bf k}_{0}}\tilde{\rho}_{\mathrm{A},{\bf k}_0} &  &  & \textit{\huge{O}}  \\
     & W_{{\bf k}_{1}}\tilde{\rho}_{\mathrm{A},{\bf k}_1} &  &  \\
    &  &  \ddots &  \\
    \textit{\huge{O}}&  & & W_{{\bf k}_{N/2-1}}\tilde{\rho}_{\mathrm{A},{\bf k}_{N/2-1}} 
    \end{bmatrix},
    \label{eq:rho_A_block_diagonal}
\end{align}
where $\{ {\bf k}_i \}$ is a set of non-equivalent wavevectors, and $\tilde{\rho}_{\mathrm{A},{\bf k}_i}$ is the nonzero $j_{\mathrm{max}}\times j_{\mathrm{max}}$ block of $\hat{\rho}_{\mathrm{A},{\bf k}_i}$:
\begin{align}
  \hat{\rho}_{\mathrm{A},{\bf k}_i} \equiv
  \begin{bmatrix}
     O& & & & \\ 
     & \ddots & & \textit{\huge{O}}& \\
    & &\tilde{\rho}_{\mathrm{A},{\bf k}_i} & & \\
    & \textit{\huge{O}}& & \ddots & \\
    & & & & O 
  \end{bmatrix}.
\end{align}
The eigenvalues of $\hat{\rho}_{\mathrm{A}}$ are therefore given by
$\{W_{{\bf k}}\lambda_{\mathrm{A},{\bf k}}\}$,
where $\lambda_{\mathrm{A},{\bf k}}$ are the eigenvalues of $\tilde{\rho}_{\mathrm{A},{\bf k}}$.
$S_{\mathrm{bi}}$ can be expressed as 
\begin{align}
  S_{\mathrm{bi}} 
  &= -\sum_{{\bf k}} \sum_{\lambda_{\mathrm{A},{\bf k}}} (W_{{\bf k}}\lambda_{\mathrm{A},{\bf k}})\ln (W_{{\bf k}}\lambda_{\mathrm{A},{\bf k}}) \nonumber \\
  &=  \sum_{{\bf k}} W_{{\bf k}} \left(
  -\sum_{\lambda_{\mathrm{A},{\bf k}}} \lambda_{\mathrm{A},{\bf k}} \ln \lambda_{\mathrm{A},{\bf k}}
  \right) -\sum_{{\bf k}} W_{{\bf k}} \ln W_{{\bf k}} 
  \nonumber \\
  &= \sum_{{\bf k}} W_{{\bf k}}S_{\mathrm{bi},{\bf k}} + S_k.
  \label{eq:Sbi_decomposition_derivation}
\end{align}
This yields the decomposition in Eq.~(\ref{eq:Sbi_decomposition}).